\pdfoutput=1

\documentclass[twocolumn,floatfix,twocolappendix]{aastex631}

\usepackage{graphicx}    

\usepackage{dcolumn}     
\usepackage{float} 

\usepackage{mathtools}
\usepackage{amsmath} 
\usepackage{amssymb} 
\usepackage{bm}          

\usepackage{tasks}
\usepackage{enumitem} 
\usepackage{multirow}

\usepackage{txfonts}

\usepackage{booktabs} 

\usepackage{cleveref}

\usepackage[english]{babel}
\usepackage[expansion=all,babel]{microtype}

\usepackage[flushmargin]{footmisc}

\newcommand{\sat}[1]{{#1}_{\mathrm{sat}}}
\newcommand{\sym}[1]{{#1}_{\mathrm{sym}}}
\newcommand{\asym}[1]{{#1}_{\mathrm{asym}}}
\newcommand{\eff}[1]{{#1}_{\mathrm{eff}}}
\newcommand{\cs}{c_\mathrm{s}}

\newcommand{\mn}{m_{\mathrm{eff;\, n}}^{\mathrm {PNM}}}
\newcommand{\mN}{m_{\mathrm{eff;\, N}}^{\mathrm {SNM}}}
\newcommand{\epnm}{\left(E/A\right)_{\mathrm{PNM}}}
\newcommand{\esnm}{\left(E/A\right)_{\mathrm{SNM}}}
\newcommand{\Yp}{Y_{\mathrm{p}}}

\newcommand{\ns}{0.16~{\rm fm}^{-3}}

\usepackage[normalem]{ulem}

\newcommand{\be}{\begin{equation}}
\newcommand{\ee}{\end{equation}}
\newcommand{\ba}{\begin{eqnarray}}
\newcommand{\ea}{\end{eqnarray}}
   
\newcommand{\Msun}{M_{\odot}}
\makeatletter
	\newcommand{\vast}{\bBigg@{2.85}}
\makeatother

\newcommand{\eq}[1]{Eq.~\eqref{#1}}

\submitjournal{ApJ}

\shorttitle{Neutron Stars Equations of State with Skyrme} 
\shortauthors{Beznogov \& Raduta}

\begin{document}

\title{Bayesian Survey of the Dense Matter Equation of State built upon Skyrme effective interactions}

\author[0000-0002-7326-7270]{Mikhail V. Beznogov}
\affiliation{National Institute for Physics and Nuclear Engineering (IFIN-HH), RO-077125 Bucharest, Romania}
\email{mikhail.beznogov@nipne.ro}

\author[0000-0001-8421-2040]{Adriana R. Raduta}
\affiliation{National Institute for Physics and Nuclear Engineering (IFIN-HH), RO-077125 Bucharest, Romania}
\email{araduta@nipne.ro}

\date{\today}


\begin{abstract}
The non-relativistic model of nuclear matter with zero-range Skyrme interactions is employed within a Bayesian approach in order to study the behavior of neutron stars (NSs) equation of state (EOS). A minimal number of constraints from nuclear physics and ab initio calculations of pure neutron matter (PNM) are imposed together with causality and a lower limit on the maximum mass of NS to all our models. Our key result is that accounting for correlations among the values that the energy per neutron in PNM takes at various densities, and that are typically disregarded, efficiently constrains the behavior of the EOS at high densities. A series of global NS properties, e.g., maximum mass, central density of the maximum mass configuration, minimum NS mass that allows for direct URCA, radii of intermediate and massive NS, appear to be correlated with the value of effective neutron mass in PNM at 0.16 fm$^{-3}$. Together with similar studies in the literature our work contributes to a better understanding of the NS EOS as well as its link with the properties of dense nuclear matter.
\end{abstract}



\section{Introduction}
\label{sec:Intro}

Hydrostatic equilibrium states of cold mature neutron stars (NSs) are determined by a one parameter equation of state (EOS) that relates pressure ($P$) to energy density ($e$). This means that any astrophysical measurement of a NS structure-related parameter, e.g., mass, radius, tidal deformability, moment of inertia, oscillation modes frequencies, probes in one way or another the EOS. Inference of the NS EOS from astrophysical observations is, nevertheless, not an easy task. First of all, only a few such measurements are available so far. Next, depending on the NS mass and the \emph{unknown} EOS, different density domains are explored throughout different stars. Then, every global parameter depends in a specific and convoluted way on the behavior the EOS has over a certain density domain. Finally, over limited density domains, different particle compositions may provide similar $P(e)$ dependencies.

Indeed, in spite of considerable effort, the NS EOS is still largely unknown \citep{Oertel_RMP_2017,Margueron_PRC_2018a,Burgio_PPNP_2021}. In addition to limited astrophysics constraints, this situation is also due to limited constraints from nuclear physics. Information from nuclear physics experiments mainly concerns a narrow density domain around nuclear saturation density $\sat{n} \approx \ns \approx 2.7 \times 10^{14}~{\rm g/cm^3}$ and matter with low isospin asymmetry, that is with similar numbers of neutrons and protons, and is customarily translated into values of nuclear empirical  parameters (NEPs). Complementary information on the density behavior of matter with extreme isospin asymmetries is provided by ab initio calculations that employ accurate nucleon-nucleon potentials and sophisticated many-body techniques. Nevertheless, so far these calculations are successful in efficiently constraining only the behavior of pure neutron matter (PNM) up to densities of the order of $\sat{n}$ \citep{Drischler_PRC_2016,Drischler_PRL_2019}.

From the astrophysical point of view, a tremendous progress was done in the last decade. As such, NS masses in the range $0.8 \lesssim M/\Msun \lesssim 2.1$~\citep{Demorest_Nature_2010, Antoniadis2013, Arzoumanian_ApJSS_2018, Cromartie2020, Fonseca_2021, Doroshenko_Nature_2022, Brodie_2023} have been measured with unprecedented precision. Measurements of massive NSs are informative of the NS EOS stiffness and of utmost relevance for the onset of non-nucleonic particle degrees of freedom \citep{Oertel_RMP_2017, Sedrakian_PPNP_2023}. At the other extremity, NS masses lower than $1~\Msun$ challenge our understanding of NS formation in core-collapse supernovae \citep{Suwa_MNRAS_2018}. This is, in particular, the case of the NS in the supernova remnant (SNR) HESS J1731-347, whose mass was recently estimated to be $M=0.77^{+0.20}_{-0.17}~\Msun$ (or $M=0.83^{+0.17}_{-0.13}~\Msun$) \citep{Doroshenko_Nature_2022, Brodie_2023}. Moreover, its very low radius $R=10.4^{+0.86}_{-0.78}~{\rm km}$ (or $R=11.25^{+0.53}_{-0.37}~{\rm km}$) \citep{Doroshenko_Nature_2022, Brodie_2023} triggers questions about the composition of low mass NS cores or the behavior of NS EOS at low densities. In relation with the first aspect, the occurrence of quark matter \citep{DiClemente_2022} and $\Delta$-resonances \citep{Li_PLB_2023} has been speculated. According to \cite{Li_PLB_2023}, the particularly compact configuration could be alternatively explained by low values of the slope of the symmetry energy. Detection of gravitational waves (GW) emitted from the coalescence of two NS in the GW170817 \citep{Abbott_PRL119_161101, Abbott_ApJ2017ApJ_L12, Abbott_PRX_2019} event or a NS-black hole binary in the GW190425 \citep{Abbott_ApJL_2020} have allowed the first inferences of NS tidal deformabilities. The relatively low masses of the NSs involved in the GW170817 event, with an estimated total mass $M_\mathrm{T}=2.73^{+0.04}_{-0.01}~\Msun$ and a mass ratio $0.73 \leq q=M_2/M_1 \leq 1$, makes that the GW signal mainly probed the intermediate density domain of the EOS.  
Observations of associated electromagnetic near-infrared event AT2017gfo \citep{Tanvir_ApJLett_2017}, gamma ray burst GRB170817A \citep{vonKienlin_2017, Savchenko_2017a,Savchenko_2017b} and optical \citep{Soares-Santos_ApJL_2017} counterparts of GW170817 have been, in turn, used to provide an estimate of the maximum mass of non-rotating cold configurations, $M_{\rm G}^* \approx 2.3~\Msun$ \citep{Margalit_17, Rezzolla_2018, Ruiz2018, Shibata_2019a}. Finally, pulse-profile modeling of X-ray data obtained by NASA’s Neutron Star Interior Composition Explorer (NICER) mission allowed simultaneous masses and radii estimates for the canonical mass pulsar PSR J0030+0451 \citep{Miller_2019,Riley_2019} and the massive millisecond pulsar J0740+6620 \citep{Miller_may2021, Riley_may2021}. In spite of the still large uncertainties on radii, of the order of a few km, these measurements can rule out a number of phenomenological EOSs.

Constraints from multimessenger astrophysics of NSs on the dense matter EOS are best exploited by statistical analyses that allow for systematic exploration of the broad parameter spaces. Many such studies have been performed in the last few years. Most of them employ parametrized EOSs models, e.g., piece-wise polytropes, spectral parameterizations and models of the speed of sound, and focus on the sensitivity of posterior distributions on the prior distributions; narrowing down of the parameter space upon progressive incorporation of constraints; dependence of the posterior distributions on the EOS model; potential benefit from measurements of extra NS observables like the moment of inertia or the gravitational binding energy; the advantage of accounting for the full posterior probability distribution of measurements rather than ``cuts''; the role that different kind of measurements play on different density ranges; compatibility between various constraints. Examples in this sense are offered by \cite{Fasano_PRL_2019, Raaijmakers_ApJ_2020,  Miller_ApJ_2020, Dietrich_Science_2020, Essik_PRD_2020, Miller_may2021, Raaijmakers_may2021}. 

Motivated by the urge for accounting for properties of nuclear matter (NM), an increasing number of works started to use phenomenological energy density functionals \citep{Lim_EPJA_2019, Zhang_ApJ_2019, Ferreira_PRD_2020, Guven_PRC_2020, Ferreira_PRD_2021, Malik_ApJ_2022, Ghosh_EPJA_2022,Patra_PRD_2022,Beznogov_PRC_2023,Malik_PRD_2023,Papakonstantinou_2023,Huang_2023}. The obvious advantage of using phenomenological models consists in having access to matter composition. Models based on the Taylor expansion of the energy per nucleon in terms of deviation from saturation and isospin symmetry \citep{Zhang_ApJ_2019, Ferreira_PRD_2020, Ferreira_PRD_2021, Patra_PRD_2022}, agnostic metamodels \citep{Guven_PRC_2020}, mean field models with Skyrme effective interactions \citep{Papakonstantinou_2023} as well as models that employ a $\chi$EFT inspired expansion in terms of neutron and proton Fermi momenta \citep{Lim_EPJA_2019} are non-relativistic. As such, causality has to be enforced up to the central density of the maximum mass configuration. Models by \cite{Malik_ApJ_2022, Ghosh_EPJA_2022, Beznogov_PRC_2023, Malik_PRD_2023, Huang_2023, Papakonstantinou_2023} rely on different formulations of the covariant density functional (CDF) theory of dense matter. Results of \cite{Malik_ApJ_2022, Malik_PRD_2023} -- who employ density dependent and nonlinear meson interactions, respectively -- indicate that NS EOS, properties, and composition depend on the EOS model and, in the case of the paper by \cite{Malik_PRD_2023} also on the assumed strength of the nonlinear scalar vector field contribution. Comparison between posterior distributions of \cite{Malik_ApJ_2022} and of \cite{Beznogov_PRC_2023} shows that even small technicalities in determining the saturation density, which enters the definition of the density dependent coupling constants, leads to small discrepancies between the results. Finally, \cite{Beznogov_PRC_2023} show that constraints on the low density behavior of PNM act differently when conditions are posed on the energy per particle or, alternatively, pressure and that the way in which the EOS is constrained impacts on the correlations between properties of NM on the one hand and properties of NS on the other hand.

The aim of the present paper is to provide better insight into the equation of state of cold catalyzed dense matter. To this end, the non-relativistic mean field approach of nuclear matter with Skyrme effective interactions is employed. This theoretical framework has been intensively used for the study of atomic nuclei but, to our knowledge, not yet utilized in a \emph{full} Bayesian inference of the NS EOS. EOS models are built by considering constraints from nuclear physics cast in terms of values of NEPs; the density dependence of the energy per neutron in PNM as calculated by the chiral effective field theory ($\chi$EFT); nucleon and neutron Landau effective masses in symmetric nuclear matter (SNM) and PNM, respectively, at the density of $\ns$, as calculated by $\chi$EFT; a lower bound on the maximum mass of a NS as well as compliance with causality at the central density of the most massive NS configuration. 

There are several reasons why we opted for this set of constraints. First, it corresponds to the strategy adopted in \citep{Malik_ApJ_2022,Malik_PRD_2023,Beznogov_PRC_2023}, meaning that the model dependence of the results can be highlighted by comparing the predictions made by these various studies. Second, by giving more weight to nuclear physics constraints, it offers a perspective complementary to that offered by studies which account for existing observational and/or potential future measurements of radii, tidal deformabilities, moments of inertia, etc. Finally, it keeps the results free from statistical and systematical uncertainties that affect some of the recent and very sophisticated astrophysical measurements on NSs. For instance, the tidal deformability extracted from GW170817~\citep{Abbott_PRL119_161101,Abbott_PRL_121,Abbott_PRX_2019} depends, among other, on whether the same EOS is used for each component star; the assumed EOS model; the spin ranges; the waveform model. Then, according to \cite{Vinciguerra_2023}, posteriors from NICER are multi-modal and subtle changes in the setup make the inference procedure prefer one mode or another, which translates into different constraints on masses and radii.

Similarly to our previous work \citep{Beznogov_PRC_2023}, different sets of constraints are examined here. Comparison between the thus obtained posterior distributions allows to judge the effectiveness of each of these constraints as well as their compatibility. Compliance with NICER and GW170817 constraints is checked a posteriori.

A purely nucleonic composition is hypothesized for the baryonic component.

The paper is organized as follows. Sec.~\ref{sec:Model} gives a brief overview of the non-relativistic mean field model with standard Skyrme interactions. Analytic expressions for several quantities, including NEPs up to the fourth order, are also provided. Basic information on the set of constraints posed to our models and the Bayesian setup are presented in Sec.~\ref{sec:Bayes}. Posterior distributions of NEPs and selected key properties of NS obtained for different sets of constraints together with some correlations between them are presented and analyzed in Sec. \ref{sec:Results}. The conclusions are drawn in Sec.~\ref{sec:Concl}. The sensitivity of the posterior distributions to the parametrization of the model and to the prior distributions is addressed in Appendix~\ref{App:MCMC}. 


\section{The model}
\label{sec:Model}

The standard Skyrme forces, that we use in this paper, have the form
\begin{eqnarray}
	V\left({\bf r}_1, {\bf r}_2 \right)&=&t_0 \left(1+x_0 P_{\sigma} \right) \delta\left( {\bf r}\right) \nonumber \\
	&+&\frac{t_1}2  \left(1+x_1 P_{\sigma} \right) \left[{\bf k}'^2 \delta\left( {\bf r}\right)
	+\delta\left( {\bf r}\right) {\bf k}^2 \right] \nonumber \\
	&+& t_2 \left(1+x_2 P_{\sigma} \right) {\bf k}' \cdot \delta\left( {\bf r}\right) {\bf k}\nonumber \\
	&+& \frac{t_3}6 \left(1+x_3 P_{\sigma} \right) \left[n \left({\bf R} \right)\right]^{\sigma} \delta\left( {\bf r}\right) \nonumber \\
	&+& i W_0 \left( \boldsymbol{\sigma}_1 + \boldsymbol{\sigma}_2\right) \cdot \left[ {\bf k}' \times \delta\left({\bf r}\right) {\bf k} \right],
\end{eqnarray}
with ${\bf r}={\bf r}_1-{\bf r}_2$, ${\bf R}=\left({\bf r}_1+{\bf r}_2\right)/2$; ${\bf k}=\left(\boldsymbol{\nabla}_1- \boldsymbol{\nabla}_2\right)/2i$ is the relative momentum operator acting on the right and ${\bf k}'$ is its conjugate acting on the left; $P_{\sigma}=\left( 1+ \boldsymbol{\sigma}_1 \cdot \boldsymbol{\sigma}_2\right)/2$ is the two body spin-exchange operator; $n({\bf r})=n_\mathrm{n}({\bf r})+n_\mathrm{p}({\bf r})$ is the total local density; $n_i({\bf r})$ with $i=\mathrm{n},\mathrm{p}$ are the neutron and proton local densities. This standard form of the Skyrme interaction has ten parameters. Eight correspond to $x_i$ and $t_i$ with $i=0,1,2,3$, one corresponds to the power $\sigma$ of the density dependence and one measures the strength $W_0$ of the zero-range spin-orbit term.

Within the independent particle approximation the total energy of a system of nucleons corresponds to the average value of the system's Hamiltonian
$\hat H$:
\begin{equation}
	\langle \psi | \hat H | \psi \rangle=\int {\cal H} \left({\bf r} \right) d {\bf r},
\end{equation}
where ${\cal H}\left({\bf r} \right)$ represents the local energy density. In the case of NM, that is homogeneous, spin-saturated and with no Coulomb interaction, the energy density ${\cal H}$ consists of four terms:
\begin{equation}
	{\cal H}=k+h_0+h_3+\eff{h},
	\label{eq:h}
\end{equation}
where $k$ is the kinetic energy term, $h_0$ is a density-independent two-body term, $h_3$ is a density-dependent term and $\eff{h}$ is a momentum-dependent term. Their expression are the following
\begin{eqnarray}
	k&=&\frac{\hbar^2}{2m} \tau,
	\label{eq:k} \\
	h_0&=&C_0 n^2+D_0 n_3^2,
	\label{eq:h0} \\
	h_3&=&C_3 n^{\sigma+2}+D_3 n^{\sigma} n_3^2,
	\label{eq:h3} \\
	\eff{h}&=&\eff{C} n \tau+\eff{D} n_3 \tau_3,
	\label{eq:heff}
\end{eqnarray}
where $n=n_\mathrm{n}+n_\mathrm{p}$ and $n_3=n_\mathrm{n}-n_\mathrm{p}$ stand for the isoscalar and isovector particle number densities; $\tau=\tau_\mathrm{n}+\tau_\mathrm{p}$ and $\tau_3=\tau_\mathrm{n}-\tau_\mathrm{p}$ denote the isoscalar and isovector densities of kinetic energy; $2/m=1/m_\mathrm{n}+1/m_\mathrm{p}$, where $m_i$ with $i=\mathrm{n},\mathrm{p}$ denotes the bare mass of nucleons. 

The kinetic energy in \eq{eq:k} takes the same form as in the Fermi gas model for noninteracting fermions.  The density behavior of the potential energy terms, Eqs.~\eqref{eq:h0}, \eqref{eq:h3} and \eqref{eq:heff}, is governed by the coefficients $C_0$, $D_0$, $C_3$, $D_3$, $\eff{C}$ and $\eff{D}$.  Each of these coefficients can be expressed in terms of traditional Skyrme parameters via a linear expression \citep{Ducoin_NPA_2006}:
\begin{eqnarray}
	C_0 &=& 3t_0/8, \nonumber\\
	D_0 &=& -t_0\left( 2 x_0+1\right)/8, \nonumber\\
	C_3 &=& t_3/16, \nonumber\\
	D_3 &=& -t_3 \left( 2 x_3+1\right)/48, \nonumber\\
	\eff{C} &=& \left[ 3 t_1+t_2 \left( 4 x_2+5\right)\right]/16, \nonumber\\
	\eff{D} &=& \left[t_2 \left( 2 x_2+1\right) -t_1 \left(2 x_1+1 \right)\right]/16.
	\label{eq:C0D0etc.}
\end{eqnarray}
Eqs.~\eqref{eq:h0}--\eqref{eq:heff} show that in the particular case of NM, the number of independent Skyrme parameters is reduced to seven. For the limiting cases of SNM and PNM the dependence is reduced further. Specifically, the energy density of SNM only depends on $\sigma$, $C_0$, $C_3$ and $\eff{C}$ while that of PNM is governed by $\sigma$, $\left(C_0+D_0\right)$, $\left(C_3+D_3\right)$ and $\left(\eff{C}+\eff{D}\right)$. The isovector and isoscalar sectors are not independent. We note that for $\sigma=0$ and $\sigma=2/3$ the functional dependence on density in \eq{eq:h} gets simplified. Indeed, for the first value \eq{eq:h3} has the same dependence as \eq{eq:h0} while for the latter value \eq{eq:h3} has the same dependence as \eq{eq:heff}. Throughout this paper, effective interactions are expressed in terms of $C_0$, $D_0$, $C_3$, $D_3$, $\eff{C}$, $\eff{D}$ and $\sigma$. This results in a parameter space of lower dimension with respect to the one associated with the traditional coefficients.

In the following we provide expressions for basic thermodynamic quantities and NEPs.

From the thermodynamic definition of pressure, $P(n,\delta)=n^2 \partial \left({\cal H}/n \right)/\partial n|_{\delta}$, where $\delta=n_3/n$ denotes the isospin asymmetry, one obtains
\begin{equation}
	P= \frac23  k+h_0+\left( \sigma+1\right) h_3+\frac53 \eff{h}.
	\label{eq:P}
\end{equation}
From \eq{eq:P} it is straightforward to compute the saturation density corresponding to an arbitrary value of $\delta$ by imposing $P(\sat{n}^{\delta})=0$ as well as to compute the value of energy per particle at the saturation, $\sat{E}(\delta)={\cal H}(n,\delta)/n|_{n=\sat{n}^{\delta}}$. We remind that saturation points exist only for $|\delta|\leq \Delta$, where $\Delta$ is an effective interaction dependent value. 

EOSs generated by various sets of parameters of the effective interaction are customarily analyzed in terms of the values NEPs have at zero temperature. There are two possible ways to achieve this goal. The first one consists in Taylor expanding the energy per particle $E/A={\cal H}/n$ of matter with the particle density $n$ and isospin asymmetry $\delta$  with respect to the deviation $\mathcal{X}=\left(n-\sat{n}^{\delta} \right)/3 \sat{n}^{\delta}$ from the saturation density $\sat{n}^{\delta}$ \citep{Dutra_PRC_2012},
\begin{equation}
	E\left(n,\delta\right)/A=\sum_{i=0,1,2,...} \frac1{i!} \sat{X}^{\delta;\,i}\, \mathcal{X}^i,
	\label{eq:Taylor_chi}
\end{equation}
with
\begin{equation}
	\sat{X}^{\delta;\,i}=3^i \left(\sat{n}^{\delta}\right)^i \left. \left( \frac{\partial^i e/n}{\partial n^i}\right) \right|_{n=\sat{n}^{\delta}}.
\end{equation}

The most important coefficients of this series are the low order ones. As any other quantity pertaining to cold matter, these coefficients can be calculated analytically as a function of Skyrme parameters. The corresponding expressions are provided in Appendix~\ref{App:NEPs}.

The second possibility is to express $E/A$ as a power series in $\delta$ \citep{Chen_PRC_2009}
\begin{equation}
	E\left(n,\delta\right)/A=E_0(n,0)+\delta^2 E_{\mathrm{sym};\,2}(n,0)+\delta^4 E_{\mathrm{sym};\,4}(n,0)+...,
	\label{eq:Taylor_delta}
\end{equation}
and then further Taylor expand each coefficient in terms of the deviation $x=\left(n-\sat{n}^{0} \right)/3 \sat{n}^{0}$ from the saturation density of SNM,
\begin{equation}
	E_0(n,0)=\sum_{i=0,1,2,...} \frac1{i!} \sat{X}^{0;\,i} x^i,
\end{equation}
\begin{equation}
	E_{\mathrm{sym};\,k}(n,0)=\sum_{j=0,1,2,...} \frac1{j!} X_{\mathrm{sym};\,k}^{0;\,j} x^j,~ k=2,4,...
\end{equation}
where $\sat{n}^{0}=\sat{n}^{\delta=0}$, $\sat{X}^{0;\,i}=\sat{X}^{\delta=0;\,i}$ and $X_{\mathrm{sym};\,k}^{0;\,j}= \left. \left(\partial^j E_{\mathrm{sym};\,k}(n,0)/\partial x^{j} \right) \right|_{n=\sat{n}^0}$.

The lowest order coefficient in $\delta^2$ present in \eq{eq:Taylor_delta} is called symmetry energy and its value at $\sat{n}^0$ is denoted by $\sym{J}$. The expressions of $E_{\mathrm{sym};\,2}$ as well as those corresponding to its
slope $\sym{L}$, curvature $\sym{K}$, skewness $\sym{Q}$ and kurtosis $\sym{Z}$ are offered in Appendix~\ref{App:NEPs}.

The asymmetry energy, $E_{\mathrm{asym}}$, is defined as the per-nucleon cost of converting SNM ($\delta=0$) in PNM ($\delta=1$),
\begin{equation}
	E_{\mathrm{asym}}(n)=E(n, \delta=1)/A-E(n, \delta=0)/A,
	\label{eq:Easym}
\end{equation}
and can further be expressed as
\begin{eqnarray}
	E_{\mathrm{asym}}(n) &=&\frac{3\hbar^2}{10 m} \left( 3 \pi^2\right)^{2/3} \left(1-\frac{1}{2^{2/3}} \right) n^{2/3} +D_0 n +D_3 n^{\sigma+1}  \nonumber \\
	&+& \left[\eff{C}  \left(1-\frac{1}{2^{2/3}} \right)+\eff{D}\right] \frac35 \left( 3 \pi^2\right)^{2/3} n^{5/3}.
	\label{eq:Easymdetailed}  
\end{eqnarray}
Numerical comparison between \eq{eq:Easymdetailed} and \eq{eq:Esym2} reveals that, for densities lower than $\sat{n}$, $E_{\mathrm{asym}} (n) \approx E_{\mathrm{sym};\,2} (n)$. Generally speaking, $E_{\mathrm{asym}}(n) \to E_{\mathrm{sym;\,2}}(n)$ whenever ${\cal O}(\delta^4) \to 0$.

For the sake of completeness we also provide expressions for the Landau effective masses. They write:
\begin{equation}
	\frac{\hbar^2}{2 m_{\mathrm{eff};\,i}}=\frac{\hbar^2}{2 m_i}+\eff{C} n \pm \eff{D} n_3,
	\label{eq:meff}
\end{equation}
where ``+'' (``--'') corresponds to neutrons (protons). The conditions to have $0 \leq  m_{\mathrm{eff};\,i}/m_i \leq 1$ lead to $\eff{C} \geq 0$ and $|\eff{D}| \leq \eff{C}$.


\section{The Bayesian setup}
\label{sec:Bayes}

In this section we discuss the various constraints posed to our EOS models and the procedure adopted to explore the parameter space.


\subsection{Constraints}
\label{ssec:Constraints}

\renewcommand{\arraystretch}{1.1}
\setlength{\tabcolsep}{8pt}
\begin{table}
	\caption{Constraints posed on EOS models.
		$\sat{E}$ and $\sat{K}$ represent the energy per particle and compression modulus of symmetric saturated matter with the density $\sat{n}$; $\sym{J}$ stays for the symmetry energy at saturation;
		$\left(E/A\right)_i$ with $i=1,2,3,4$ stand for the energy per particle of PNM at the densities of 0.04, 0.08, 0.12 and 0.16 $\mathrm{fm}^{-3}$;
		$M_{\mathrm{G}}^*$ represents the maximum gravitational mass of a NS; $\cs^{*2}$ stands for the speed of sound squared  at a density equal to the central density of the maximum mass configuration.
		$\mn$ ($\mN$) denotes the Landau effective mass of the neutron  (nucleon) in PNM (SNM) at $\ns$.
		For most quantities we provide the median value and the standard deviation;
		for $M_{\mathrm{G}}^*$ and $\cs^{*2}$, we specify the threshold values.
		$m_\mathrm{n}$ and $m_\mathrm{N}$ represent the neutron and nucleon bare masses, respectively.}
	\centering
	\begin{tabular}{ccccc}
		\toprule
		\toprule
		Quantity             & Units              & Value   & Std. deviation & Ref. \\
		\midrule
		$n_{\mathrm{sat}}$   & $\mathrm{fm}^{-3}$ & 0.16 	& 0.004          & 1 \\
		$E_{\mathrm{sat}}$   & MeV 		          & $-15.9$ & 0.2 	         & 1 \\
		$K_{\mathrm{sat}}$   & MeV 		          & 240 	& 30 	         & 1 \\ 
		$J_{\mathrm{sym}}$   & MeV 		          & 30.8    & 1.6            & 1  \\
		$\left(E/A\right)_1$ & MeV 	              & 6.165   & 0.059          & 2 \\
		$\left(E/A\right)_2$ & MeV 	              & 9.212   & 0.226          & 2 \\
		$\left(E/A\right)_3$ & MeV 	              & 12.356  & 0.512          & 2 \\       
		$\left(E/A\right)_4$ & MeV 	              & 15.877  & 0.872          & 2 \\       
		$\mn$                & $m_\mathrm{n}$     & 0.880   & 0.026          & 2 \\       
		$\mN$                & $m_\mathrm{N}$     & 0.638   & 0.013          & 2 \\       
		$M_{\mathrm{G}}^*$   & $\Msun$ 	          & $>2.0$  &  ---           & 3     \\
		$\cs^{*2}$           & $c^2$              & $<1$    & ---            & --- \\
		\bottomrule
		\bottomrule
	\end{tabular}
	\label{tab:constraints}
{\raggedright \textbf{References.} \citet{Margueron_PRC_2018a}; (2) \citet{Drischler_PRC_2021}; (3) \citet{Fonseca_2021}.\par}
\end{table}
\setlength{\tabcolsep}{2.0pt}
\renewcommand{\arraystretch}{1.0}

\renewcommand{\arraystretch}{1.05}
\setlength{\tabcolsep}{1.9pt}
\begin{table}
	\centering
	\caption{Sets of constraints, other than those on $\sat{n}$, $\sat{E}$, $\sat{K}$, $\sym{J}$, $M_{\mathrm{G}}^*$ and $\cs^{*2}$, considered in our runs. For the meaning of $(E/A)_i$ with $i=1,2,3,4$, $\mN$ and $\mn$, see the caption of Table~\ref{tab:constraints}.}
	\begin{tabular}{cccccccc}
		\toprule
		\toprule
		Run   & $\left(E/A\right)_1$ & $\left(E/A\right)_2$ & $\left(E/A\right)_3$ & $\left(E/A\right)_4$ & correl.    & $\mN$ & $\mn$ \\
		\midrule
		0     & --                   & \checkmark           &  \checkmark         & \checkmark           & \checkmark &  -- & --    \\
		1     & --                   & \checkmark           &  \checkmark         & \checkmark           & --         &  -- & --    \\
		2     & --                   & \checkmark           &  \checkmark         & \checkmark           & --         &  \checkmark & --    \\
		3     & --                   & \checkmark           &  \checkmark         & \checkmark           & --         &  -- &  \checkmark   \\
		4     & \checkmark           & \checkmark           &  \checkmark         & \checkmark           & --         &  -- & --    \\
		\bottomrule
		\bottomrule
	\end{tabular}
	\label{tab:runs}
\end{table}
\setlength{\tabcolsep}{2.0pt}
\renewcommand{\arraystretch}{1.0}

A minimum number of constraints is imposed to models built in this work. They correspond to the behavior of NM around $\left(n = \sat{n}, \delta=0 \right)$; the energy per neutron in PNM [$\epnm$] at densities lower or equal to the saturation density; nucleon (neutron) Landau effective mass in SNM (PNM) at $n=\ns$; lower bound on the maximum gravitational mass of a NS and causality of NS matter. For values, see Table~\ref{tab:constraints}.

The NM-related conditions refer to the values the best known NEPs, $\sat{n}$, $\sat{E}$, $\sym{J}$ and $\sat{K}$, have. For each of the considered NEPs, median values and standard deviations (SDs) are fixed to the values obtained by \cite{Margueron_PRC_2018a} based on a collection of 35 standard Skyrme interactions that are frequently employed in the literature (SDs are rounded up).
This choice ensures that the domains of values used as a reference correspond to \emph{well-behaved} energy density functionals \emph{identical} to those assumed here. We remind at this point that the values obtained for NEPs depend on both the assumed energy density functional and the analyses itself, for a discussion see \citep{Oertel_RMP_2017,Margueron_PRC_2018a}. Out for the four NEPs in Table~\ref{tab:constraints}, $\sat{K}$ is the least controlled one \citep{Oertel_RMP_2017}.

The density behavior of PNM is controlled by the values the energy per particle has in a minimum of three and a maximum of four points with densities in the range $0.04 ~\mathrm{fm}^{-3} \leq n \leq 0.16 ~\mathrm{fm}^{-3}$. For all these energies we use the values provided by \cite{Drischler_PRC_2021}, where results of $\chi$EFT calculations corresponding to six Hamiltonians \citep{Drischler_PRC_2016} are reported. The nucleon-nucleon (NN) interactions computed at N$^3$LO are supplemented by three-nucleon (3N) interactions computed at N$^2$LO. The spread in the energy per particle obtained from these interactions can serve as an uncertainty estimate. More importantly, access to the individual energy per particle values provided by each Hamiltonian, rather than to the values of the means and SDs, allows to calculate correlations between the values corresponding to different densities, see Sec.~\ref{ssec:Likelihood}.

Some of the models built in this paper also impose constraints on the values the nucleon (neutron) effective mass has in SNM (PNM) at $n=\ns$. Similarly to the density behavior of the energy per particle in PNM, the values obtained by \cite{Drischler_PRC_2021} are used. Due to the limited flexibility of Landau effective masses as functions of density in models with standard Skyrme interactions (see \eq{eq:meff} and Sec.~\ref{ssec:NM}), constraints on $\mN$ and $\mn$ are not imposed over a wider density domain despite being available from \cite{Drischler_PRC_2021}.

The fact that the values used as a reference for NEPs on the one hand and $\epnm$, $\mn$ and $\mN$ on the other hand stem from different sources and have been derived within dissimilar theoretical frameworks might, in principle, raise some tension. In particular, we note that \cite{Margueron_PRC_2018a} provides for $\mN/m_{\mathrm N}$ at saturation a value, $0.77 \pm 0.14$, that is by $20\%$ higher that the one of \cite{Drischler_PRC_2021}. Besides, according to Fig.~3 by \cite{Drischler_PRC_2021}, $\sat{n}=0.163~\mathrm{fm}^{-3}$ and $\sat{E}=-15.07~\mathrm{MeV}$. The first value agrees, within $1 \sigma$, with the value in Table~IV of \cite{Margueron_PRC_2018a}, while the latter is by $6\%$ lower than the value put forward in Table~IV of \cite{Margueron_PRC_2018a}. The compatibility between constraints derived from phenomenological and ab initio models will be addressed elsewhere.

NS matter is minimally constrained. First, we ask that any model complies with the $2~\Msun$ lower limit on the maximum NS mass. Second, we ask that causality is fulfilled at the density corresponding to the central density of the maximum mass configuration.

We note that each of the NM constraints in Table~\ref{tab:constraints} acts on a different number of input parameters of the model.
As such, constraints on $\sat{n}$, $\sat{E}$, $\sat{K}$ get translated into constraints on
$\sigma$, $C_0$, $C_3$ and $\eff{C}$;
constraints on $\epnm$ lead to constraints on $\sigma$,
$\left(C_0+D_0\right)$, $\left(C_3+D_3\right)$, $\left(\eff{C}+\eff{D}\right)$;
constraints on $\mN$ lead to constraints on $\eff{C}$;
constraints on $\mn$ lead to constraints on $\left(\eff{C}+\eff{D}\right)$;
constraints on $\sym{J}$ act on $D_0$, $D_3$, $\eff{C}$, $\eff{D}$, $\sigma$; notice that
$\sym{J}$ depends on $C_0$ and $C_3$ through $\sat{n}$.

In a ``deterministic approach'' four sharp constraints on $\left(E/A \right)_{PNM}$ would translate into well defined values for $\sigma$, $\left(C_0+D_0\right)$, $\left(C_3+D_3\right)$, $\left(\eff{C}+\eff{D}\right)$; should an extra sharp constraint on $\mn$ be posed at any non-vanishing density, with a high probability, no compatible model will be found. If this is the case, values for the (linear combinations of) effective interaction parameters can be determined through a $\chi^2$-fitting procedure. This is exactly the procedure used to construct phenomenological effective interactions based on reproduction of nuclear physics observables. As a matter of fact, in most of these cases the number of constraints exceeds the number of the degrees of freedom of the interaction. The fact that, with the exception of run 4, the energy of PNM is constrained in three density values only, confers flexibility to our sets of models; extra flexibility comes from the fact that, instead of providing sharp values, we allow the quantities on which constraints are imposed to span some domains.

Finally, note that constraints imposed on NS properties cannot be connected in a straightforward manner with either parameter of the effective interaction, the reason being that the value that each of these quantity takes depends on the behavior of matter over wide domains of density and proton fraction.


\subsection{Likelihood}
\label{ssec:Likelihood}

Similarly to the situation we have considered previously \citep{Beznogov_PRC_2023}, only the lower limit on $M_\mathrm{G}^*$ and the upper limit on $c_\mathrm{s}$ can be regarded as ``purely experimental'' information. All other constraints are obtained employing sophisticated theoretical modeling. As such, a question of their independence/uncorrelatedness/compatibility arises. We will not go into details here. Note that the topic of independence/uncorrelatedness was discussed at lengths by \cite{Beznogov_PRC_2023} (cf. also Supplemental Materials of \cite{Drischler_PRC_2021}). Some issues regarding compatibility will be addressed in Sec.~\ref{sec:Results}. Here we only focus on the differences compared to the situation of \cite{Beznogov_PRC_2023}.

First, since we have employed for the energy per particle of PNM a set of data \citep{Drischler_PRC_2016,Drischler_PRC_2021}, where values are provided for each of the six considered Hamiltonians, it is possible to take into account correlations when constraining $\epnm$ at different values of density. Following the procedure described by \cite{Drischler_PRC_2021} in the Supplemental Materials, we can compute the covariance matrix as follows:
\begin{align}
	\mathrm{cov}_\mathit{ij} = \frac{1}{5}\sum_{k=1}^6 {\cal E}_k(n_i) {\cal E}_k(n_j) - \frac{1}{5} \sum_{k=1}^6 {\cal E}_k(n_i)\, \frac{1}{5} \sum_{k=1}^6 {\cal E}_k(n_j).
	\label{eq:Covariance}
\end{align}
Here $n_i$ and $n_j$ are the values of density at which we compute the correlations between the values of the energy per particle ${\cal E}=E/A$ and $k$ runs over the six Hamiltonians investigated by \cite{Drischler_PRC_2021}. Considering that the mean is also estimated from the same sample, we have introduced Bessel's correction (i.e., dived by 5 instead of 6) to get an unbiased estimate of the covariance matrix. 

The log-likelihood function for runs 1--4 has the ``typical'' form for uncorrelated constraints \citep{Beznogov_PRC_2023}:
\begin{align}
	\log \mathcal{L}_q \propto -\chi_q^2 = -\frac{1}{2} \sum_{i=1}^{N_q}  \left(\frac{d_i - \xi_i(\mathbf{\Theta}) } {\mathcal{Z}_i} \right)^2
	\label{eq:Chi2-NoCorr}
\end{align}
where $q$ labels the run, $N_q$ is the number of constraints for that run except for $M_\mathrm{G}^*$ and $\cs^{*2}$ constraints, $d_i$ and $\mathcal{Z}_i$ stand for the constraint and its SD as mentioned in Table~\ref{tab:constraints}; $\xi_i(\mathbf{\Theta})$ corresponds to the value the model defined by the parameter set $\mathbf{\Theta}$ provides for the quantity $i$. However, for run 0 the log-likelihood reads: 
\begin{align}
	\begin{split}
		&\log \mathcal{L}_0 \propto -\chi_0^2 =\\ 
		&-\frac{1}{2} \sum_{i=1}^{N_0}  \left(\frac{d_i - \xi_i(\mathbf{\Theta})} {\mathcal{Z}_i} \right)^2 
		-\frac{1}{2}\sum_{i=1}^3 \sum_{j=1}^3 \left(\mathrm{cov}^{-1}\right)_\mathit{ij} \delta {\cal E}_i \delta {\cal E}_j,
	\end{split}
	\label{eq:Chi2-Corr}
\end{align}
where first term corresponds to the constraints that we always treat as uncorrelated (e.g., $\sat{n}$, $\sat{E}$, $\sat{K}$, $\sym{J}$) and the second term deals with energy per particle constraints that we consider to be correlated in run 0; $\delta {\cal E}$ stands for the difference between the value of the constraint and the value provided by the model.

In Eqs.~\eqref{eq:Chi2-NoCorr} and \eqref{eq:Chi2-Corr} we have discarded the normalization factors as they are not relevant for sampling from the posterior distributions (as long as they do not depend on the input parameters, which they do not in our case). In the same way as in \citep{Beznogov_PRC_2023}, our primary method of Bayesian inference was affine invariant Markov Chain Monte Carlo (MCMC). See the details of implementation, employed software, analysis of convergence and of the effects of the prior in Appendix~\ref{App:MCMC}. 

Second, now we have two ``impenetrable wall'' constraints. One is the lower limit  on $M_\mathrm{G}^*$ and the other is the upper limit on $\cs^{*2}$. They are dealt with in a similar fashion to \citep{Beznogov_PRC_2023}, i.e., if one or another (or both) is violated, the log-likelihood function is assigned a large negative value ($-10^{10}$). This insures that such a point is never accepted into the Markov chain. By employing a similar rejection procedure, we have also discarded all configurations where NS matter was thermodynamically unstable, i.e., where pressure was either negative (at some densities) or decreasing with density.


\subsection{Prior}
\label{ssec:Prior}

The last crucial ingredient of any Bayesian-like analysis is the prior. For technical reason described in Appendix~\ref{App:MCMC}, we have employed ``mixed'' input parameters, i.e. instead of using all seven effective interaction parameters ($C_0$, $D_0$, $C_3$, $D_3$, $\eff{C}$, $\eff{D}$ and $\sigma$) as the input ones, we have actually used three NEPs ($\sat{n}$, $\sat{E}$ and $\sym{J}$) and four effective interaction parameters ($D_3$, $\eff{C}$, $\eff{D}$ and $\sigma$) as the input. The remaining parameters of the effective interaction, $C_3$, $C_0$ and $D_0$, were computed according to
\onecolumngrid
\begin{align}
	\begin{split}
		&C_3=\frac{1}{\sigma {\sat{n}}^{\sigma+1}}\left[-\sat{E} + \frac{\hbar^2}{10m} \left( \frac{3 \pi^2 }{2}\right)^{2/3}\sat{n}^{2/3} -\frac{2}{5} \eff{C} \left( \frac{3 \pi^2}{2} \right)^{2/3} \sat{n}^{5/3}
		\right],\\
		&C_0=-\frac{1}{{\sat{n}}} \left[\frac{\hbar^2}{5 m}  \left( \frac{3 \pi^2}{2} \right)^{2/3}\sat{n}^{2/3}
		+\left(\sigma+1 \right) C_3 {\sat{n}}^{\sigma+1} + \eff{C}  \left( \frac{3 \pi^2}{2} \right)^{2/3}\sat{n}^{5/3}\right], \\
		&D_0=\frac{\sym{J}-J_k}{\sat{n}}-D_3 {\sat{n}}^{\sigma}-\left( \frac{\eff{C}}{3}+\eff{D}\right)
		\left( \frac{3 \pi^2}{2} \right)^{2/3} {\sat{n}}^{2/3},
	\end{split}
	\label{eq:re-param}
\end{align}
with $J_k=\hbar^2/6m \left(3\pi^2/2 \right)^{2/3}\sat{n}^{2/3}$. 
\twocolumngrid

\renewcommand{\arraystretch}{1.05}
\setlength{\tabcolsep}{8.0pt}
\begin{table}
    \vspace{0.5cm}
	\caption{Domains of the priors.}
	\centering
	\begin{tabular}{cccc}
		\toprule
		\toprule
		Parameter      & Units                          & Min.              & Max.             \\
		\midrule
		$\sat{n}$      & $\mathrm{fm^{-3}}$             & 0.14              & 0.18             \\
		$\sat{E}$      & MeV                            & $-16.9$           & $-14.9$          \\
		$\sym{J}$      & MeV                            & 22.8              & 38.8             \\
		$D_3$          & MeV $\mathrm{fm^{3+3\sigma}}$  & $-6.0\times 10^3$ & $3.0\times 10^3$ \\
		$\eff{C}$      & MeV $\mathrm{fm^{5}}$          & 0.0               & 900            \\
		$\eff{D}$      & MeV $\mathrm{fm^5}$            & $-900$            & 900            \\
		$\sigma$       &  --                            & 0.01              & 1.1              \\
		\bottomrule
		\bottomrule
	\end{tabular}
	\label{tab:Prior}
\end{table}
\setlength{\tabcolsep}{2.0pt}
\renewcommand{\arraystretch}{1.0}

The domains of the priors are presented in Table~\ref{tab:Prior}. For the three NEPs inputs, the domains cover 10 SDs range symmetrically around the median value (cf. Table~\ref{tab:constraints}). For $D_3$, the domain was chosen somewhat arbitrarily based on trial and error procedure. The domains of $\eff{C}$ and $\eff{D}$ take into account that the effective masses should lie within $0.02 \lesssim m_{\mathrm{eff}}/m_{\mathrm{N}} \leq 1$, with the lower limit chosen arbitrarily. The domain we choose for $\sigma$ is slightly larger than the one span by the ensemble of Skyrme interactions, see Table~B1 in \cite{Dutra_PhD}, where parameters of the effective interactions are compiled for 240 models. Indeed, according to Table~B1, $0.08 \leq \sigma \leq 1.00$. Thus, our choice encompasses all behaviors allowed by standard parametrizations. See also  Appendix~\ref{App:MCMC} regarding the importance of $\sigma$ domain range.

We have chosen uniform (uninformative) priors in the ranges provided in Table~\ref{tab:Prior}. Notice, however, that the constraint $|\eff{D}| \leq \eff{C}$ (see Sec.~\ref{sec:Model}) means that the actual prior distribution of $\eff{C}$ and $\eff{D}$ will not be uniform after this constraint is enforced. Also notice that uniform priors in $\sat{n}$, $\sat{E}$, $\sym{J}$, $D_3$, $\eff{C}$, $\eff{D}$ and $\sigma$ are not equivalent to uniform priors in the seven effective interaction parameters. See details in Appendix~\ref{App:MCMC}.  


\section{Results}
\label{sec:Results}

In this section we analyze posterior probability density distributions of various quantities pertaining to SNM, PNM, NS matter and NS properties, as obtained for each of the five runs. For the sake of the comparison, we also show the envelope of 75 standard Skyrme models from \cite{Dutra_PRC_2012}, which satisfy the conditions: 
i) NS matter is thermodynamically stable, i.e., $P_\mathrm{NS}$ is positive and non-decreasing as a function of density,
ii) $M_{\mathrm{G}}^*/\Msun \geq 2$, iii) $\cs^2(n_\mathrm{c}^*)\leq 1$, iv) $22.8~\mathrm{MeV} \leq \sym{J} \leq 38.8~\mathrm{MeV}$. The latter condition corresponds to a dispersion within 5~SDs from the central value
reported in Table~\ref{tab:constraints}. Hereafter, we refer to these 75 models as \emph{Comparison Set} models. For each situation we discuss the impact of each type of constraints. Finally, a number of correlations among NM and NS parameters is discussed.

Before analyzing the results it is worth mentioning that different runs in this work have different ``fitting'' quality. While all the runs of \cite{Beznogov_PRC_2023} have very similar values of $\chi^2$ for their respective maximum a posteriori (MAP) points, in the present work values of $\chi^2$ for MAP points are noticeably different. They are:
\begin{tasks}[style=itemize](2)
        \task run~0: $\chi^2_\mathrm{MAP} = 2.63$
        \task run~1: $\chi^2_\mathrm{MAP} = 0.07$
        \task run~2: $\chi^2_\mathrm{MAP} = 0.06$
        \task run~3: $\chi^2_\mathrm{MAP} = 1.48$
        \task run~4: $\chi^2_\mathrm{MAP} = 0.34$
\end{tasks}
This is our first indication that standard Skyrme model might not be flexible enough to account for all of the imposed constraints. Further evidence is discussed in the following subsections.

\subsection{Nuclear matter}
\label{ssec:NM}

\begin{figure*}
	\centering
	\includegraphics[]{"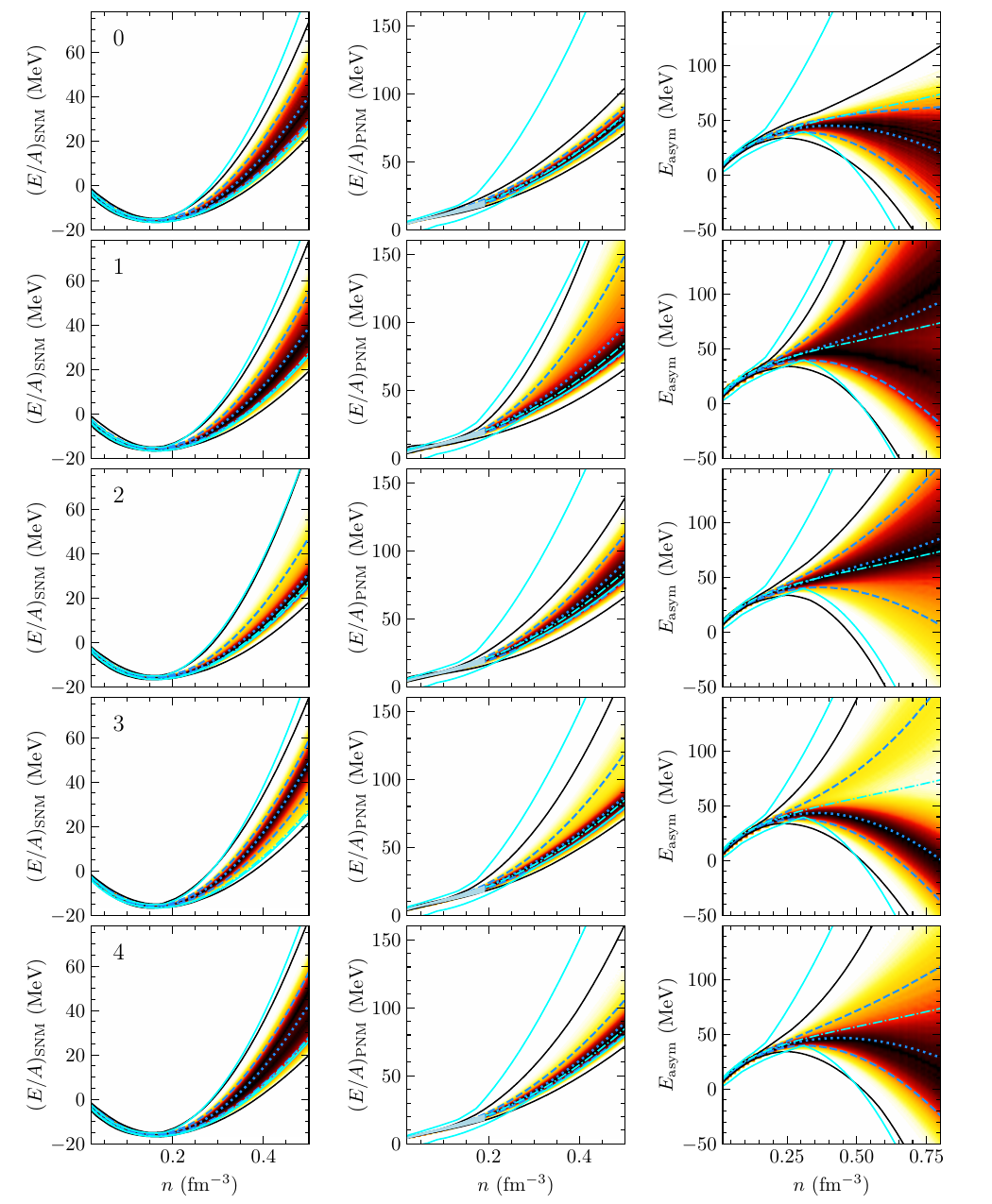"}
	\caption{Conditional probability density (also known as curve density) plots corresponding to the density dependence of the energy per nucleon in SNM (left column) and PNM (middle column) and asymmetry energy, \eq{eq:Easym}, (right column). Results of various runs are illustrated on subsequent rows, as indicated on the left panels. Curve density is indicated by colors: dark red (light yellow) corresponds to large (low) densities. Dotted and dashed dodger blue curves demonstrate the median and 90\% CR, respectively. Solid black curves mark the envelope of the bunch of models associated with each run. Solid cyan curves mark the envelope of Comparison Set models, cyan dot-dashed curves show their median, see text for details. Light blue shadowed regions mark the uncertainty domain for $\epnm$, as computed by \cite{Drischler_PRC_2021}.
	}
	\label{Fig:CD_NM}
\end{figure*}

\begin{figure*}
	\centering
	\includegraphics[]{"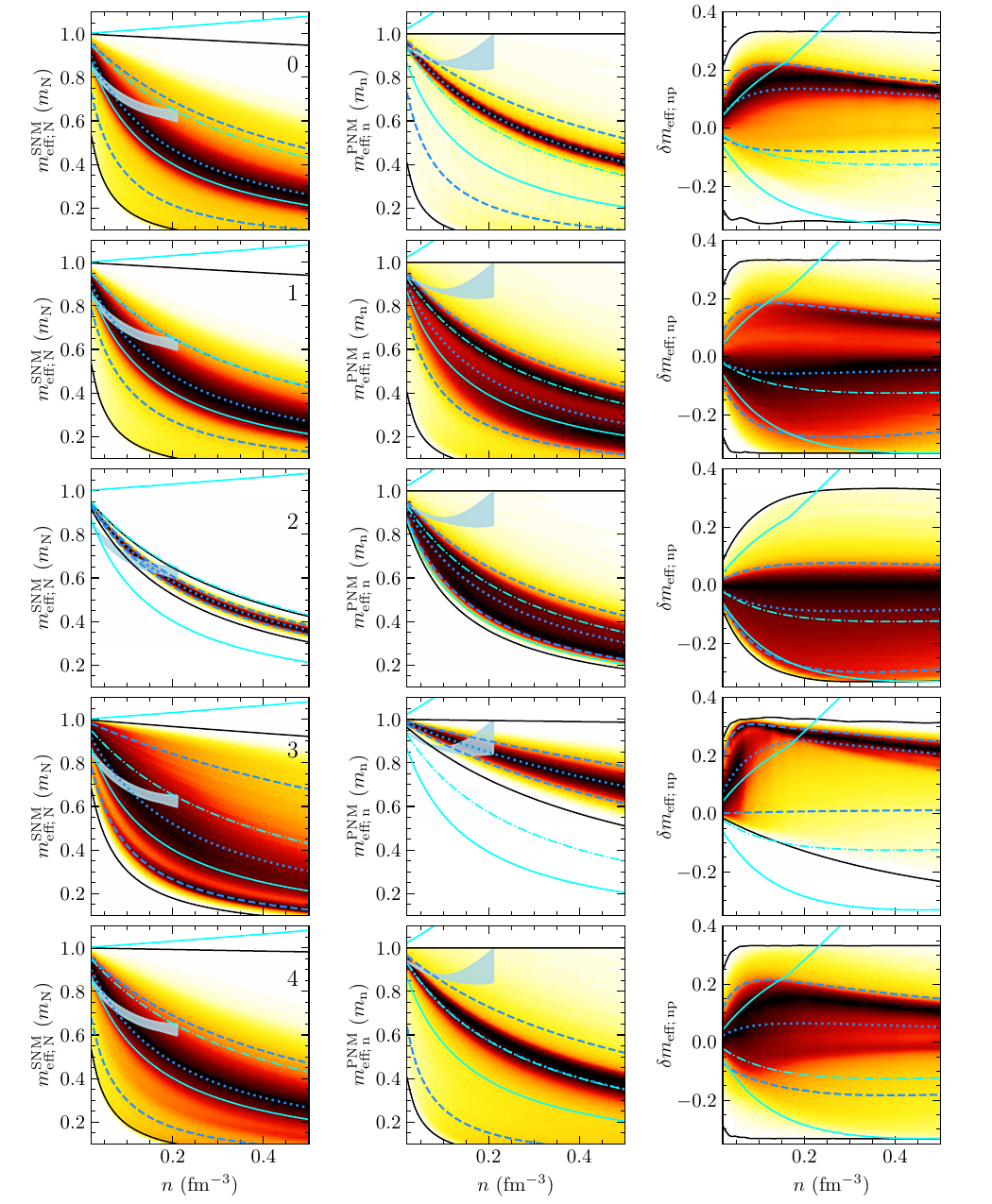"}
	\caption{Conditional probability density (also known as curve density) plots corresponding to the density dependence of the nucleon effective mass in saturated SNM (left column), neutron effective mass (middle column) and neutron-proton effective mass splitting in neutron-rich matter with $Y_\mathrm{p}$=0.2 (right column). Results of various runs are illustrated on subsequent rows, as indicated on the left panels. The color map shows the curve density. Dotted and dashed dodger blue curves demonstrate the median and 90\% CR, respectively. Solid black curves mark the envelope of the bunch of models associated with each run. Solid cyan curves mark the envelope of Comparison Set models, cyan dot-dashed curves show their median, see text for details. Light blue shadowed regions mark uncertainty domains as computed by \cite{Drischler_PRC_2021}.
	}
	\label{Fig:CD_meff}
\end{figure*}

\begin{figure*}
	\centering
	\includegraphics[]{"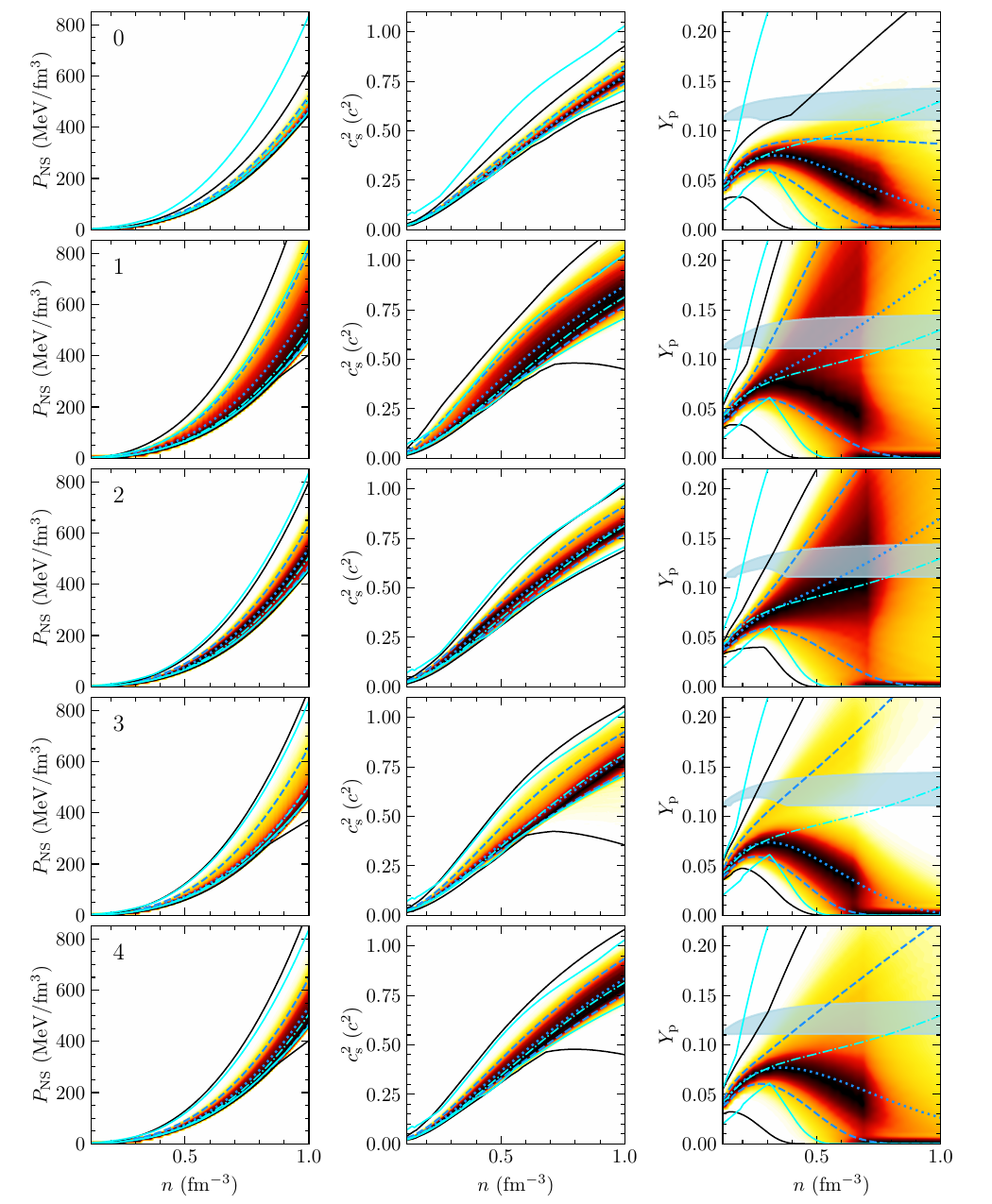"}
	\caption{Conditional probability density (also known as curve density) plots corresponding  to the density dependence of pressure (left column), speed of sound squared (middle column) and proton fraction (right column) in NS matter. Results of various runs are illustrated on subsequent rows, as indicated on the left panels. The color map shows the curve density. Dotted and dashed dodger blue curves demonstrate the median and 90\% CR, respectively. Solid black curves mark the envelope of the bunch of models associated with each run. Solid cyan curves mark the envelope of the Comparison Set models, cyan dot-dashed curves show their median. The light blue domains on the right panels correspond to the direct URCA threshold, as predicted by our models; see text for details.
	}
	\label{Fig:CD_NS}
\end{figure*}

The density dependence of the energy per nucleon in SNM and PNM is investigated in Fig.~\ref{Fig:CD_NM} (left and middle columns). Results corresponding to different runs in Table~\ref{tab:runs} are reported on subsequent rows.  The considered density domain is $0.02 \leq n \leq 0.5~{\rm fm}^{-3}$. This and the following (Figs.~\ref{Fig:CD_meff} and \ref{Fig:CD_NS}) conditional probability density plots of the posterior distributions are normalized such that for any given fixed value of $n$ the total probability (of a considered function of $n$ to have any value) is 1.0. The colormap is normalized accordingly, i.e., black color corresponds to the maximum probability density at each fixed $n$ (unlike the joint probability density plots where the maximum of the colormap would correspond to the maximum of probability density in the whole 2D domain, not at each cross-section). In this and the following figures, CR stands for credible region.

Regarding the behavior of SNM we note that:\\
(i) different runs feature rather similar dispersions between $\esnm$ curves; this is understandable given that the behavior of the isoscalar channel is constrained by conditions imposed on $\sat{n}$, $\sat{E}$ and $\sat{K}$ and those are common for all runs; curves in each family start to diverge at a density slightly larger than $\sat{n}$; this is a remainder that constraints at $\sat{n}$ are not sufficient to control the high density behavior of the EOS,\\
(ii) one more constraint posed on $\epnm$ over $n \leq \sat{n}$ does not seem to impact the scattering of curves over the complementary domain $n>\sat{n}$; indeed, the difference between the dispersion of curves in runs~1 and 4, which contains one extra constraint with respect to run~1, is small,\\
(iii) accounting for correlations among $\epnm$ over $n \leq \sat{n}$ slightly reduces the dispersion of curves over the complementary domain of density (run~0 vs. run~1),\\
(iv) implementation of the constraint on $\mN$ slightly broadens the uncertainty band of $\esnm$ over $n>\sat{n}$ (run~2 vs. run~1),
while no such effect is obtained when the constraint is imposed on $\mn$ (run~3 vs. run~1), \\
(v) the envelopes corresponding to each of our runs and the one that corresponds to the Comparison Set models are of similar width; overall, models built in this work prefer slightly lower values than those of Comparison Set models,\\
(vi) the median curves of runs~0 and~4 are close to the median curve in run~1; the median curve in run~3 (run~2) explores larger (lower) values than those of the median curve in run 1.

Now, let us consider the consequences that various constraints have on $\epnm$. The following comments are in order:\\
(i) run 0, which accounts for correlations among $\epnm$ at 0.08, 0.12 and $0.16~\mathrm{fm}^{-3}$, provides by far the smallest dispersion of curves; this means that accounting for correlations at $n \leq \sat{n}$ is very efficient in controlling the behavior of PNM even at densities much larger than $\sat{n}$, \\
(ii) any extra constraint imposed on the behavior of PNM at low densities results in reducing the curve dispersion over $n>\sat{n}$, see the uncertainty bands in runs~4 and 1,\\
(iii) constraints on the effective mass at saturation are efficient in reducing the dispersion (runs~2 and 3 vs. run~1); constraints on $\mN$ are more efficient than those on $\mn$; the reason is that conditions on $\mN$ control $\eff{C}$, while those on $\mn$ control $\left( \eff{C}+\eff{D} \right)$; together with $\left( C_0+D_0 \right)$ and $\left( C_3+D_3 \right)$, $\left( \eff{C}+\eff{D} \right)$ is also controlled by the conditions on $(E/A)_{i}$, though in a loose way (see Fig.~\ref{Fig:Hist_ModelParam}),\\
(iv) the widths of the envelopes that correspond to all runs except run~1 are much lower than the width of the envelope associated with the Comparison Set models; this means that the models in each of our sets manifest less variety in the density dependence of the symmetry energy; the situation can be attributed to the constraints we have imposed on $(E/A)_{i}$ and that are absent in many standard Skyrme interactions,\\
(v) the largest (lowest) value of the median curve at $n= 0.5~\mathrm{fm}^{-3}$ corresponds to run 1 (run 0), that is, to the run with the largest (lowest) dispersion, \\
(vi) the constraints we have imposed play a much larger role on $(E/A)(n)$ in PNM than on $(E/A)(n)$ in SNM;
this is easy to understand given that most of these constraints are imposed on the behavior of PNM, \\
(vii) over $n \lesssim \ns$, the posterior distributions of all our runs are in good accord with the results from $\chi$EFT \citep{Drischler_PRC_2021}, illustrated by a light blue band. 

The right column in Fig.~\ref{Fig:CD_NM} addresses the density dependence of $\asym{E}$, \eq{eq:Easym}. Largely different behaviors of PNM allowed by models belonging to each run are translated into largely different density dependencies of the asymmetry energy. In particular, runs 0 and 1, which provide the most and the least constrained behavior of $\epnm(n)$ (see the middle column), respectively, manifest the smallest and largest dispersion of $\asym{E}(n)$ at $n \geq \sat{n}$. It is worth noting that all our runs accommodate models where $\asym{E}$ increases with density along with models that show the opposite behavior. Models in the second category allow for negative values of $\asym{E}$ at densities exceeding a certain value. The Comparison Set models manifest the same variety of behaviors. The only cases where the median curve increases with density correspond to run~1 and run~2.

Recently, constraints on the symmetry energy and symmetry pressure ($\sym{P}=n^2 \partial \sym{E}/\partial n$) at around $(2/3) \sat{n}$ have become available. They have been obtained either from homogeneous sets of measurements/calculations or compilations of data that span a range of densities. Examples are provided by $P_{\mathrm{sym};0.1}=2.38 \pm 0.75~\mathrm{MeV/fm^3}$, extracted from the neutron skin thickness of $^{208}\mathrm {Pb}$~\citep{PREXII,Reed_PRL_2021}; $P_{\mathrm{sym};0.1}=1.8 \pm 0.2~\mathrm{MeV/fm^3}$, inferred from a collection of nuclear structure and heavy ion collision data and astrophysical observations of NSs~\citep{Lynch_PLB_2022}; $E_{\mathrm{sym};0.1}=25.6^{+1.4}_{-1.3}~\mathrm{MeV}$, obtained from a correlation between a linear combination of chemical potentials in nuclei neighboring shell closures and the symmetry energy at $(2/3) \sat{n}$~\citep{Qiu_PLB_2024}. A $\chi$EFT $E_{\mathrm{asym};0.1}$ value can be computed from the energies per nucleon in SNM and PNM provided in \citep{Drischler_PRC_2021} and amounts to $23.65 \pm 0.37~\mathrm{MeV}$. The values of $E_{\mathrm{sym};0.1}$ and $P_{\mathrm{sym};0.1}$ corresponding to each of our runs are reported in Table~\ref{tab:Posteriors}. In all cases, the medians and SDs of these quantities are slightly smaller than those in \citep{Lynch_PLB_2022,Qiu_PLB_2024}. The relative similitude with $P_{\mathrm{sym};0.1}$ in \citep{Lynch_PLB_2022} can be attributed to the equivalence of the yet different constraints. The relative similitude with $E_{\mathrm{sym};0.1}$ in \citep{Qiu_PLB_2024} is most probably attributable to our usage of realistic constraints on $\sym{J}$ and $E/A$ in PNM. Our $E_{\mathrm{sym};0.1}$-values are in excellent agreement with $E_{\mathrm{asym};0.1}$ extracted from \citep{Drischler_PRC_2021}; this can be seen as an a posteriori proof of effectiveness of constraints on $E/A$ in PNM. Our $P_{\mathrm{sym};0.1}$-values are much lower than those in \citep{Reed_PRL_2021}. This, however, should not come as a surprise given that the values of $\sym{J}=38.1 \pm 4.7~\mathrm{MeV}$ and $\sym{L}=106 \pm 37~\mathrm{MeV}$ in \citep{Reed_PRL_2021} exceed considerably those in this work, see Sec.\ref{sssec:NMParam}, as well as other values extracted from nuclear physics~\citep{Oertel_RMP_2017}. 

Further insight into the properties of NM is offered in Fig.~\ref{Fig:CD_meff}, where the effective nucleon mass in SNM (left column), effective neutron mass in PNM (middle column) and neutron-proton effective mass splitting in neutron rich matter with $Y_p=0.2$ (right column) are plotted as a function of density. The results corresponding to the Comparison Set models and the predictions of $\chi$EFT calculations of \cite{Drischler_PRC_2021} are illustrated as well. We note that all models generated within all of the runs are characterized by effective masses smaller or equal to the bare mass and that they decrease or stay constant with the density. This is a direct consequence of conditions imposed on $\eff{C}$ and $\eff{C} \pm \eff{D}$ (see Sec.~\ref{sec:Model}) in order to produce this behavior. Note that this is nevertheless not the case of all standard Skyrme interactions is the literature. For instance, Ska45s20 \citep{Dutra_PRC_2012}, Sefm09 \citep{Dutra_PRC_2012}, Sefm1 \citep{Dutra_PRC_2012}, SK255 \citep{Agrawal_PRC_2003}, which are present in the Comparison Set, lead to $\mn$ increasing with density. Ska45s20 also leads to $\mN$ increasing with density. Other examples are offered by BSk1 \citep{Samyn_NPA_2002} and BSk2 \citep{Samyn_NPA_2002} for which both $\mn$ and $\mN$ augment with the density. The density behavior of the upper boundary of these models' envelope is due to Ska45s20 (Sefm1) for SNM (PNM).

The light blue shadowed areas, illustrating the predictions of microscopic calculations of \cite{Drischler_PRC_2021}, show that the decrease in $\mN(n)$ gets saturated around $0.2~{\rm fm}^{-3}$ and $\mn(n)$ is non-monotonic. As obvious from the analytic expressions in \eq{eq:meff}, the standard Skyrme effective interactions are too simple to account for such a behavior. In order for it to be described a more complex density dependence of the effective mass is needed. Bruxelles extended Skyrme models \citep{Chamel_PRC_2009} provide a possible solution and the performances of more sophisticated Skyrme interactions will be considered elsewhere.

The left column in Fig.~\ref{Fig:CD_meff} shows that, with the exception of run~2 where a constraint is posed on the value of the effective mass of nucleons in SNM at the density $\ns$, the EOS models in each run manifest a large dispersion of $\mN(n)$; the largest dispersion by 90\% CR corresponds to EOS models in run 3, where a constraint is posed on the value of neutron effective mass in PNM.
The scattering of models in run~2 is  small, which means that the constraint is efficient. Moreover, one can see that the target value of $\mN$ at $\ns$ is met rather accurately. The dispersion that characterizes runs 0, 1, 3 and 4 makes that a significant percentage of our models predict for $\mN$ values smaller than the those of the Comparison Set. This situation is attributable to the much larger variety of models explored here. We also note that absence of constraints sensitive to the value of $\mN$ in the fitting procedure of the Comparison Set models makes that some of them have $\mN$ increasing with density and exceeding the bare nucleon mass. 

The middle column in Fig.~\ref{Fig:CD_meff} shows that $\mn(n)$ is very sensitive to the way in which the constraints are incorporated. This situation is easy to anticipate. Accounting for correlations among the values of $(E/A)_{i}$ in PNM at different densities makes models gather in a relatively narrow valley though the envelope remains as large as when correlations are disregarded (run~0 vs. run~1). Constraints on effective masses at the saturation density result in the accumulation of models in a narrow band (runs~2 and 3 vs. run~1). Posterior distributions from run 2 and run 3 differ much one from the other. As expected, the effect is stronger when the constraint is imposed on the neutron effective mass in PNM than on the nucleon effective mass in SNM. The first of these constraints also leads to values of $\mn(n)$ larger than those obtained when the second constraint is imposed (run~3 vs. run~2). The fact that at $\ns$ the values of $\mn$ in run 3 are close to the target values means that the constraint is effectual. The effect obtained when an extra constraint is posed on $(E/A)$ in PNM at low densities (run~4) is weaker than the one obtained when correlations among values of $(E/A)_{i}$ are accounted for (run~0). We remind that, for obvious reasons, this was also the case of $(E/A)(n)$ in PNM, see Fig.~\ref{Fig:CD_NM} (middle column). We note that the domain of maximum curve density in run~4 has a width intermediate to widths of domains of maximum curve density in run~0 and run~1. Finally, from all runs, the upper value border of all models is characterized by $\mn(n)/m_n=1$, which corresponds to $\eff{C}=-\eff{D}$.

The right column in Fig.~\ref{Fig:CD_meff} illustrates the density dependence of $\delta m_\mathrm{eff;\,np}=m_\mathrm{eff;\,n}/m_\mathrm{n}-m_\mathrm{eff;\,p}/m_\mathrm{p}$ in neutron-rich matter with the proton fraction $Y_\mathrm{p}=n_\mathrm{p}/n=(1-\delta)/2$ equal to 0.2 (arbitrarily chosen). For all runs a wide spread of curves is observed; the values range between $-0.35$ and $0.35$. Runs~0 and 3 mostly generate positive values, which means that most models agree with the predictions of microscopic calculations that include three body forces \citep{Sammarruca_PRC_2005, Dalen_PRC_2005, Baldo_PRC_2014, Catania_PRC_2020}. We remind that all of the latter calculations provide values of neutron Landau effective mass that exceed, in neutron rich matter, the values of the proton Landau effective mass. This agreement does not come as a surprise given that run~0 accounts for correlations among values of $(E/A)$ in PNM at different densities and run~3 explicitly implements a condition on the neutron effective mass in PNM. These two runs also produce the strongest accumulation of models in a narrow domain; they correspond to the situations where the smallest dispersion of $\mn(n)$ is obtained. The median curves in these runs explore values round 0.1 (0.2) for run~0 (run~3). Run~2 mostly generates negative values; this shows that conditions posed on $\mN$ at $\ns$ play differently than those posed on $\mn$ at $\ns$. The median curve in run~2 is characterized by values of the order $-0.1$. The models in the runs~1 and 4 are more uniformly distributed around $\delta m_\mathrm{eff;\,np} = 0$; models in the run~4, which implements constraints on a wider density domain, gather in a moderately narrow band centered around 0.05 while no such behavior is shown by models in run~1.  Positive (negative) values of $\delta m_\mathrm{eff;\,np}$ are obtained by models with $\eff{D}<0$ ($\eff{D}>0$). Positive and negative values of $\delta m_\mathrm{eff;\,np}$ are obtained also by models in the Comparison Set; the median curve of this bunch of models is characterized by $\delta m_\mathrm{eff;\,np} < 0$. 


\subsection{Neutron stars matter}
\label{ssec:NS}

In this subsection we consider the role that each of the constraints we impose play on the NS EOS. NS EOSs are obtained by smoothly matching the core and the crust EOSs. For the outer and inner crusts we adopt the models by \cite{HDZ_1989} and \cite{NV_1973}, respectively. The smooth matching between crust and core EOSs is done at a density of about $\sat{n}/2$. In the leptonic sector both electrons and muons are accounted for. Modifications brought to the NS stiffness are investigated by analyzing the density dependence of pressure ($P_\mathrm{NS}$) and speed of sound squared ($\cs^2$). We remind that $\cs^2/c^2=\left( \mathrm{d}P/ \mathrm{d}e \right)_\mathrm{fr}$, where the subscript “fr” indicates that the derivatives have to be evaluated with the composition frozen. Modifications brought to the chemical composition are judged from the density dependence of the proton fraction $\Yp$. The predictions of the five sets of models considered in this paper are depicted in Fig.~\ref{Fig:CD_NS}.

The left and middle columns show that the patterns in $(E/A)(n)$ in PNM are somehow replicated in $P_\mathrm{NS}(n)$ and $\cs^2 (n)$. More precisely, the smallest and largest dispersions among the $P_\mathrm{NS}(n)$ and $\cs^2 (n)$ curves generated within one run correspond to the run~0 and run~1, respectively, which show the smallest and largest dispersions among $(E/A)(n)$ in PNM curves, see Fig.~\ref{Fig:CD_NM} (middle column). 
For the huge majority of models $\cs^2$ is a monotonic function of $n$. In this situation, models with values of the speed of sound exceeding, at high $n$-values, the speed of light should not be considered as violating causality. The situation should be rather understood as the consequence of prolonging the curves beyond the largest density value for which stable NSs configurations are obtained. Indeed, compliance with causality is imposed in our models by requiring that, for a density equal to the central density of the maximum mass NS configuration ($n_{\mathrm c}^*$), $\cs^2 \leq 1$. 
For a tiny fraction of the models the maximum value of $\cs^2$ is nevertheless reached at a density lower than $n_{\mathrm c}^*$. Some of those may violate causality. We have checked that, due to their low number, such models have no effect on the statistics.

The sudden change of slope in the lower boundary of the pressure envelope of the runs 1, 3 and 4 is attributable to the softening that the softest models in these runs suffer at high densities. This phenomenon is responsible also for the behavior of the lower boundary of the $\cs^2$ envelope that, starting from $0.6 - 0.7~{\rm fm}^{-3}$, decreases with the density, see the middle column. 
With the exception of runs 0 (the most constrained) and 1 (the least constrained), the models generated here manifest roughly the same scattering in $P_\mathrm{NS}(n)$ as the Comparison Set models, though models in the latter set do not exhibit the extra softening at high densities. As a consequence, the agreement between the values that $\cs^2$ spans in the runs 2, 3, and 4 and the values that $\cs^2$ covers in the Comparison Set holds only up to $\approx 0.7~\mathrm{fm}^{-3}$.

The widely different behaviors of $\asym{E}(n)$ in our runs lead to a large dispersion of the proton fraction $\Yp(n)$ in $\beta$-equilibrated matter. Indeed, all the five runs contain models where $\Yp$ increases steeply with the density as well as models where $\Yp$ vanishes at densities a few times larger than $\sat{n}$. Runs~1 and 2 show that the models that yield large (small) values of $\Yp$ also yield large (small) values of $P_\mathrm{NS}$, the strength of this correlation depending on the value of density. Note, nevertheless, that this correlation is not universal, as $P_\mathrm{NS}(n)$ has an explicit dependence on $\Yp(n)$, $\sym{E}(n)$ and $d\sym{E}(n)/dn$ (see Eq.~(101) of \cite{Steiner_PhysRep_2005}). Models with $\Yp$ steeply increasing with $n$ will allow the direct URCA process to operate even in low mass stars. It is know that in order for direct URCA to operate, $\Yp$ should exceed a threshold value. The latter depends on the electron ($n_\mathrm{e}$) and muon ($n_{\mu}$) densities and writes $Y_{\mathrm{p;\,DU}}=1/\left[1+\left( 1+x_\mathrm{e}^{1/3}\right)^3 \right]$, with $x_\mathrm{e}=n_\mathrm{e}/\left( n_\mathrm{e}+n_{\mu} \right)$. Models with vanishing $\Yp$ not only do not accommodate for direct URCA, but lead to NSs whose inner core is exclusively made of neutrons. The shape of the lower envelope of the Comparison Set models shows that such a behavior is not an artifact of our models. Indeed, vanishing $\Yp$ values are caused by negative values of $\asym{E}$. We note that only the models in the runs~1 and 2 have median curves that increase with the density, which corresponds to $\sym{E}$ (or $\asym{E}$) increasing with the density, see Fig.~\ref{Fig:CD_NM} (right column).

\begin{figure*}
	\centering
	\includegraphics[]{"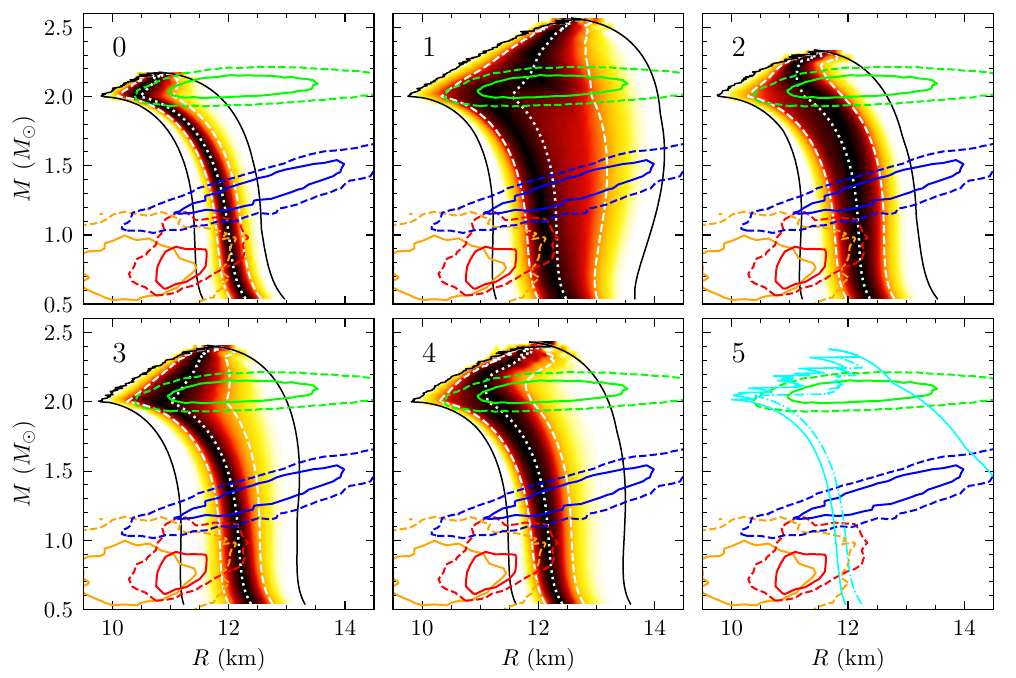"}
	\caption{Conditional probability density (also known as curves density) plots $P(R|M)$. The color map shows the curve density. Dotted and dashed light cyan curves demonstrate the median and 90\% CR, respectively. Solid black curves mark the envelope of the bunch of models associated with each run. The envelope and the median the Comparison Set models are illustrated with cyan solid and dot-dashed curves in the panel ``5''. Only stable configurations are plotted. Also depicted are joint mass and radius posterior probability density contours corresponding to PSR J0030+0451 \citep{Miller_2019} (blue contours), PSR J0740+6620 \citep{Miller_may2021} (green contours) and to the NS in the SNR HESS J1731--34 \citep{Doroshenko_Nature_2022} (orange and red contours); solid and dashed lines stand for 50\% and 90\% CR, respectively.
	}
	\label{Fig:CD_MR}
\end{figure*}
\begin{figure}
	\includegraphics[]{"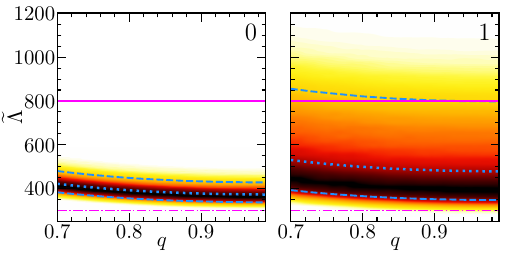"}
	\caption{Conditional probability density plots $P(\widetilde\Lambda|q)$. Dotted and dashed dodger blue curves illustrate the median and 90\% CR, respectively. Horizontal dot-dashed and solid magenta lines correspond to the median and 90\% symmetric upper bound of the $300^{+500}_{-190}$ constraint on the combined tidal deformability $\widetilde\Lambda$ from GW170817~\citep{Abbott_PRX_2019}. Left and right panels correspond to runs~0 and 1, respectively.
	}
	\label{Fig:CD_qL}
\end{figure}

$M-R$ diagrams corresponding to the models in each run are represented in Fig.~\ref{Fig:CD_MR} as conditional probability density (also known as curve density) plots $P(R|M)$. Unlike the previous conditional probability density figures, this plot requires an additional normalization. The reason is that for $M > 2~\Msun$ the number of curves is no longer constant, it decreases until there is only one curve left corresponding to the maximum of the maximum masses. Thus, we have first to normalize the 2D bins' counts corresponding to a given value of $M$ to the number of EOSs that exist for this value of $M$. Then we normalize the colormap to the maximum probability density at each fixed value of $M$ similarly to the previous conditional probability density plots (for the latter, see the explanation in the beginning of Sec.~\ref{ssec:NM}). Results corresponding to each run are depicted in a separate panel. It comes out that the run~0 and run~1, which provide the smallest and the largest dispersion among the models, populate the narrowest and the broadest domains of $M-R$ space, respectively. For instance, the difference among the largest and smallest radii of $1.4~\Msun$ stars is 1.2~km for run~0 and 3.0~km for run~1. The corresponding figures for the $2.0~\Msun$ stars are 2.0~km and 4.3~km, respectively. The lowest (largest) values of maximum gravitational mass, $M_{\mathrm{G}}^*$, are obtained for the run~0 (run~1), which shows the softest (stiffest) $P_\mathrm{NS}(n)$ dependence (see Fig.~\ref{Fig:CD_NS} left column). The runs~0 and 1 distinguish one from the other also by the sensitivity NS radii have to NS masses. Indeed, correlations between $(E/A)$ in PNM at different densities for models in the run~0 are translated in NS radii which decrease with the increase of NS masses. The lowest number of constraints implemented in the run~1 is responsible for a more ambiguous evolution of $R(M)$, especially in low mass stars. The largest radii predicted by the Comparison Set models for stars with $M/\Msun \leq 1.5$ are significantly larger than the largest radii predicted by our models. These large values are due to the models with particularly steep dependence of $\asym{E}$ as a function of $n$, see the right column of Fig. \ref{Fig:CD_NM}. The fact that this situation does not hold also in more massive stars is a nice illustration that the density dependence of symmetry energy impacts only low and intermediate mass NSs.

Joint mass and radius posterior probability density contours corresponding to PSR J0030+0451 \citep{Miller_2019}, PSR J0740+6620 \citep{Miller_may2021} and to the NS in the SNR HESS J1731--34 \citep{Doroshenko_Nature_2022} are also plotted in each of the six panels in Fig.~\ref{Fig:CD_MR} at 50\% (solid lines) and 90\% (dashed lines) CR. We present two variants of constraints for HESS J1731--34: orange contours correspond to the ``X-ray only'' data of \cite{Doroshenko_Nature_2022}, while red contours correspond to the ``full priors'' data. Note that the latter data include specific assumptions on the NS EOS \citep[see][]{Doroshenko_Nature_2022}. Those assumptions can be incompatible with our assumptions, thus rendering the comparison between our results and the ``full prior'' dataset meaningless. The obvious conclusions are that, i) in each of the five runs there are models that comply at 50\% CR with ``constraints'' from PSR J0030+0451 and PSR J0740+6620, ii) none of those also comply with any of the 50\% CR ``constraints'' from HESS J1731--34. The same holds for the Comparison Set models. These features signal either a tension between the $\chi$EFT-inspired conditions that we have imposed on $\epnm$ and HESS J1731--34, or insufficient flexibility of the density functional we have assumed. This aspect will be addressed elsewhere.

Predictions of the least and most constrained runs in what regards the combined tidal deformability $\widetilde\Lambda$ as a function of the mass ratio $q$ are confronted in Fig.~\ref{Fig:CD_qL}, where the constraints from GW170817~\citep{Abbott_PRX_2019} are also plotted. In addition to the expected shrinking of the uncertainty band upon accounting for correlations among $(E/A)_i$ in PNM, we note that median values in run~0 are somewhat lower than those in run~1. The values spanned in run~0 are slightly higher than the median value from \citep{Abbott_PRX_2019}; the 90\% CR in run~1 is very close to the 90\% upper bound from \citep{Abbott_PRX_2019}, but this is a coincidence. One can see that our calculations result in larger values of $\widetilde\Lambda$ compared to the GW data. They are, nevertheless, well within 90\% confidence interval of the observational constraints.


\subsection{Model dependence of the results}
\label{ssec:ModeDep}

The role that the different constraints play on models of NM and NSs is discussed in this subsection by analyzing 1D marginalized posterior distributions of different quantities. Considered are: parameters of the effective interactions as well as linear combinations of them; parameters of NM; energy per particle in PNM at various densities lower or equal to $\ns$; nucleon (neutron) effective mass in SNM (PNM); global properties of NS.

\subsubsection{Posterior distributions of parameters of the effective interaction}
\label{sssec:EffIntParam}

\begin{figure*}
	\centering
	\includegraphics[]{"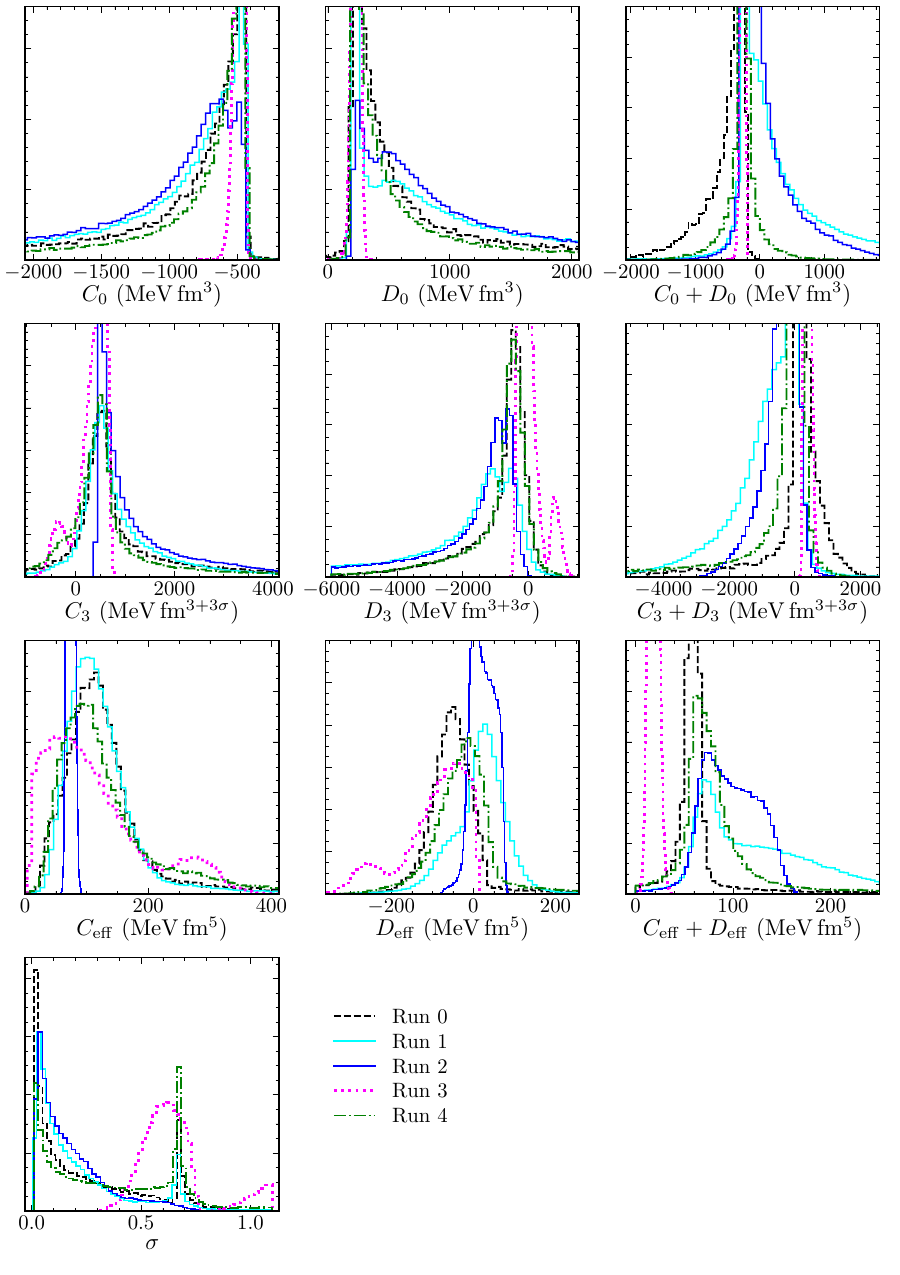"}
	\caption{Marginalized posteriors for the parameters of standard Skyrme effective interactions. Also shown are marginalized posteriors for $\left(C_0+D_0 \right)$, $\left(C_3+D_3 \right)$, $\left(\eff{C}+\eff{D} \right)$ on which constraints on $\left(E/A \right)$ in PNM and $\mn$ act. All the five runs from Table~\ref{tab:runs} are considered. X- and Y-axis ranges were chosen to increase readability. As such, some of the very wide or very narrow distributions are cut at X-axis or Y-axis limits, respectively.
	}
	\label{Fig:Hist_ModelParam}
\end{figure*}

In Sec.~\ref{sec:Bayes} we mentioned that four sharp constraints on the energy per particle in PNM at various densities would result in attributing values to $\left(C_0+D_0 \right)$, $\left(C_3+D_3 \right)$, $\left(\eff{C}+\eff{D} \right)$ and $\sigma$. The same would be the case when sharp constraints are posed on the value that the energy per particle in PNM takes for three values of the density and one sharp constraint is posed on the value that $\mn$ takes at one density. By employing a Bayesian framework, we expect to obtain posterior distributions for each of those quantities rather than single values. Marginalized posterior distributions of run~4 and run~3, which correspond to the situations described above and which are illustrated in Fig.~\ref{Fig:Hist_ModelParam}, show that the distributions of the first three parameters mentioned above are peaked. The peaks are particularly narrow for run~3. Besides, low values of $\left(\eff{C}+\eff{D} \right)$ are preferred for run~3. The posterior distribution of $\sigma$ clearly shows two peaks but their widths and localizations are different in these two runs. The peaks in run~4 are situated slightly above 0 and around 2/3, respectively. As discussed in Sec.~\ref{sec:Model}, $\sigma=0$ and $\sigma=2/3$ correspond to situations where the functional relation in \eq{eq:h} gets simplified and a certain degeneracy is introduced in both isoscalar and isovector channels. This feature makes the conditions imposed on the various quantities easier to meet, which translates into accumulation of points. The posterior distribution of $\sigma$ in run~3 peaks around $2/3$ and towards the maximum value allowed in the simulation, 1.1; also,  the values $\sigma \lesssim 0.3$ are forbidden. The preference for large values of $\sigma$ as well as the suppression of low values of $\sigma$ stem from the necessity to compensate for the low values of $\left(\eff{C}+\eff{D} \right)$, which decrease the $\eff{h}$ term of pressure, see Eq.~\eqref{eq:P}. Yet, the pressure should be high enough to comply with the $2~\Msun$ constraint. The low population of the narrow domain around $\sigma \approx 3/4$ can be regarded as the outcome of strong competition among $\sigma=2/3$ and $\sigma=1.1$.

Run~1 corresponds to a looser version of run 4. This suggests that posterior distributions of $\left(C_0+D_0 \right)$, $\left(C_3+D_3 \right)$, $\left(\eff{C}+\eff{D} \right)$ should be wider than those for run~4. Fig.~\ref{Fig:Hist_ModelParam} confirms also this expectation; in particular, the strongest widening is obtained for $\left(\eff{C}+\eff{D} \right)$. This figure also shows that the posterior distribution of $\sigma$ in run 1 is peaked at the same values as the posterior distribution for run 4, though this time preference is given to low  $\sigma$ values. This corroborates the interdependence between $\left(\eff{C}+\eff{D} \right)$ and $\sigma$ noticed in the previous paragraph.

Correlations between values taken by $\left(E/A \right)$ in PNM at different densities, accounted for in run~0, suggest that the peaks in $\left(C_0+D_0 \right)$, $\left(C_3+D_3 \right)$, $\left(\eff{C}+\eff{D} \right)$ should be narrower than those corresponding to run~1. Fig.~\ref{Fig:Hist_ModelParam} shows that this is, indeed, the case. It also shows that the distribution of $\sigma$ resembles qualitatively the distribution in run~4 and run~1. A clear preference for the lowest allowed value 0.01 is to be noticed. 

In run~2 the extra constraint is posed on $\mN$, which translates into a constraint on $\eff{C}$. The posterior distribution of $\left(\eff{C}+\eff{D} \right)$ becomes rather wide and the large value tail of $\sigma$ distribution is suppressed, which makes that the peak around $\sigma=2/3$ disappears.

Information on the posterior distributions for the $C_0$, $D_0$, $C_3$, $D_3$, $\eff{C}$, $\eff{D}$ parameters of the effective interaction is provided in the other panels. It turns out that the distributions of the $C$-parameters somewhat mirror the distributions of the $D$-parameters. The only exception from this rule corresponds to $\eff{C}$ and $\eff{D}$ in run~2, where the ``symmetry'' is ``broken'' by the constraint on $\mN$ that acts only on $\eff{C}$. In fact, run~2 is the only case where the distributions of $\eff{C}$ and (to a lesser extent) $\eff{D}$ are narrow. This coherence between $C$- and $D$-parameters obviously means that the behaviors in isovector and isoscalar channels are not independent. 

Finally, we note that, in some cases, the posterior distributions of some parameters are cut. Specifically, these are $D_3$ in runs~1 and 2 and $\sigma$ in run~3. This situation is the direct consequence of the chosen priors, see Sec.~\ref{sec:Bayes}, and suggests that slightly different distributions would be obtained if parameters were allowed to span wider domains. Note that other parameters that appear to be cut on Fig.~\ref{Fig:Hist_ModelParam} are not really cut, they just go outside the axis limits, which were chosen to increase readability.

\subsubsection{Posterior distributions of parameters of NM, effective masses and $(E/A)$ in PNM}
\label{sssec:NMParam}

\begin{figure*}
	\centering
	\includegraphics[]{"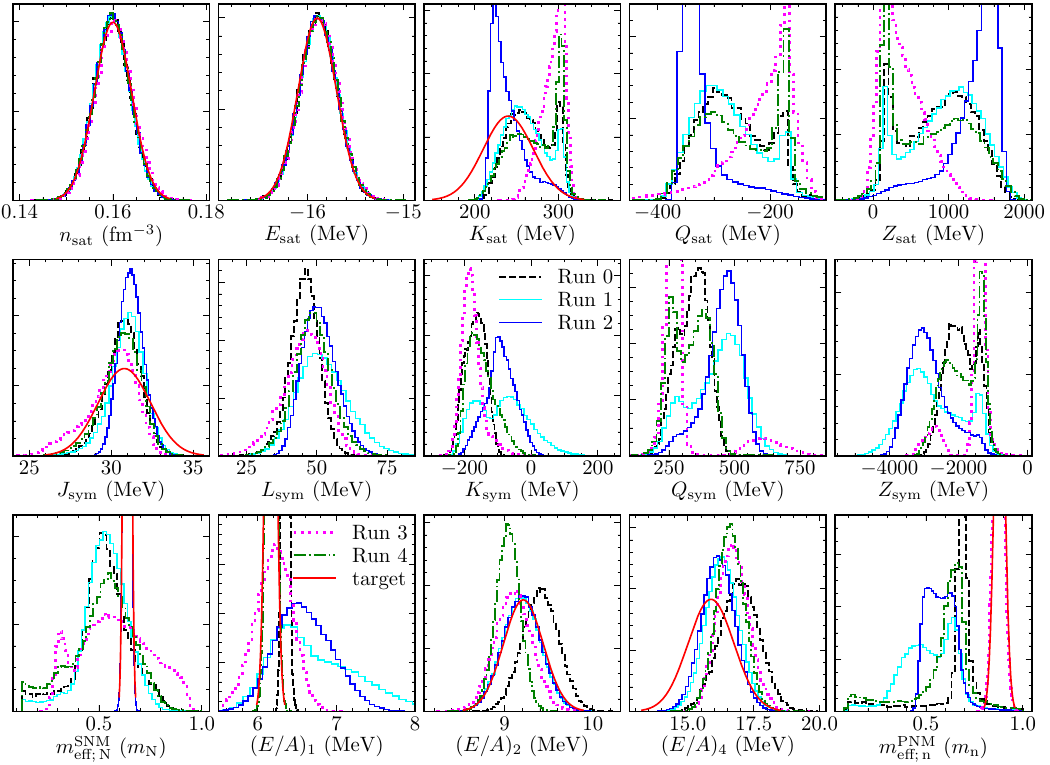"}
	\caption{Marginalized posteriors of NEPs, effective nucleon mass in SNM, effective neutron mass in PNM and energy per particle in PNM at densities between 0.04 and $\ns$. All runs are considered. When available, target distributions for constrained parameters are also plotted. Y-axis ranges have been chosen to increase readability. As such, some of the very narrow distributions are cut.
	}
	\label{Fig:Hist_NM}
\end{figure*}

In order to increase the efficiency, the MCMC exploration is done by controlling $\sat{n}$, $\sat{E}$ and $\sym{J}$, see Sec.~\ref{ssec:Likelihood} and Appendix~\ref{App:MCMC}.

The first two quantities are allowed to span narrow domains of values, see Table~\ref{tab:constraints}. Fig.~\ref{Fig:Hist_NM} shows that  marginalized posterior distributions meet very well the target distributions. The fact that distributions corresponding to different runs sit one on the top of the other is explained by the fact that the constraint on $\sat{K}$ is common to all runs and other constraints in Table~\ref{tab:constraints}, which are specific to one run or another, have limited effect on the zero order parameters in the isoscalar channel. Indeed, out of the parameters that govern the density dependence in the isoscalar channel, the constraints on $(E/A)$ in PNM only act on $\sigma$ while that on $\mN$ acts only on $\eff{C}$. The posterior distributions of $\sat{K}$ differ much from one run to another. This shows that the high order parameters in the density dependence of symmetric matter do depend on constraints on PNM as they do depend on extra constraints on SNM. This result is another manifestation of the entanglement of isoscalar and isovector sectors in Skyrme models, already visible in \eq{eq:h3}. Distributions of $\sat{K}$ in run~0, run~1 and run~4 are double peaked and qualitatively similar; each of the peaks is correlated with a peak in the posterior distribution of $\sigma$. Distributions of $\sat{K}$ in run~2 and run~3 differ from distributions of other runs and among themselves; the same was the case of posterior distributions of $\sigma$. The fact that all the posterior distributions of $\sat{K}$ miss the target distribution can be regarded as an incompatibility between different conditions that we have imposed. Patterns in the distributions of $\sat{K}$ are replicated in the distributions of higher order parameters $\sat{Q}$ and $\sat{Z}$. Distributions of these latter quantities also extend over wider domains, which shows that uncertainties in the density behavior of the EOS increase with the density. It appears that low values in $\sat{K}$ are correlated with low values in $\sat{Q}$ and high values in $\sat{Z}$. Notice that the same holds for EOSs derived within CDF theory, irrespective whether the meson couplings depend on density \citep{Malik_ApJ_2022,Beznogov_PRC_2023} or include non-linear terms \citep{Malik_PRD_2023}, which suggests that those correlations are universal.

Posterior distributions of $\sym{J}$ in runs~0, 1, 2 and 4 are bell-shaped, while the corresponding distribution in run~3, where an extra constraint was posed on $\mn$, is asymmetric; posterior distributions in runs~0, 1 and 4 are quite similar and cover the same domain of values while those for runs~2 and 3 are shifted towards larger and lower values, respectively. Runs~2 and 3 also provide the narrowest and largest distributions. Given that runs~0, 1 and 4 contain only constraints on $\epnm$, while runs~2 (3) also constrains $\mN$ ($\mn$), the result looks natural. While none of the runs is successful in reproducing the shape of the target distribution, the peak value is met with relatively high accuracy. This was not the case of the $\sat{K}$ distributions. Posterior distributions of $\sym{L}$ are bell-shaped for all runs and differ somewhat from one run to another. We note in particular that run~1, that allows for the largest values of $\asym{E}$ (see Fig.~\ref{Fig:CD_NM}), also allows for the largest values of $\sym{L}$. Discrepancies in $\sym{L}$ mean that the various sets of constraints are effectual in controlling the low density behavior of the isovector channel and that each set of constraints act differently. The situation is very different in what regards high order parameters. Indeed, the distributions of $\sym{K}$, $\sym{Q}$ and $\sym{Z}$ are wide and increasingly different from one run to the other. We note in particular that the only double peaked distribution of $\sym{K}$ corresponds to run~1, while the only single-peaked distributions of $\sym{Q}$ and $\sym{Z}$ correspond to run~2. Peaks in the posterior distributions of $\sym{Q}$ and $\sym{Z}$ appear to be correlated with the peaks in $\sigma$. Patterns in the $\sym{Q}$ distributions are to some extent replicated in the $\sym{Z}$ distributions and low values of one quantity are correlated with high values of the second. This negative correlation between $\sym{Q}$ and $\sym{Z}$ reminds the negative correlation between $\sat{Q}$ and $\sat{Z}$ discussed in the previous paragraph. The similitude between the behaviors of the third and fourth order parameters in the isoscalar and isovector sectors is striking the more if we note that negative correlations between $\sym{Q}$ and $\sym{Z}$ are spotted also in CDF models \citep{Malik_ApJ_2022,Beznogov_PRC_2023,Malik_PRD_2023} though their strength is model-dependent.

The target distributions of $\mN$ and $\mn$ at $\ns$ are met only in runs~2 and~3, where conditions on these quantities are posed. In other runs, the posterior distributions of the two effective masses show complex structures and extend over wide domains. Distributions of $\mN$ at $\ns$ in runs~0, 1 and, to a lesser extent, 4 are similar; there are no two runs that provide, for $\mn$, similar distributions. The double peaked structure of $\mN$ ($\mn$) in run~3 (run~1) obviously originates from the double peaked structure of $\eff{C}$ ($\eff{C}+\eff{D}$), see Fig.~\ref{Fig:Hist_ModelParam}. 

Small SDs of $\left(E/A\right)$ in PNM at low densities make that the corresponding constraints are very efficient. As a consequence, run~4 (runs~1, 2 and, to a lesser extent, 3) meet the target distribution at $0.04~{\rm fm}^{-3}$ ($0.08~{\rm fm}^{-3}$) while in all other cases target distributions on $\left(E/A\right)$ in PNM  are missed. The situation of $\left(E/A\right)_1$ in run~0 is interesting as even if its distribution does not agree with the target, it is still fairly close to it and of comparable width. Notice that for run~0 the value of $\left(E/A\right)$ in PNM at $0.04~\mathrm{fm}^{-3}$ is not explicitly constrained. This result is the out-turn of accounting for correlations at different (but higher) densities. We also note that the distribution of $\left(E/A\right)_1$ for run~1 is wide. This means that the uncorrelated constraints posed at higher densities are ineffective at low densities. The fact that also the distribution of $\left(E/A\right)_1$ for run~2 is wide shows that the extra constraint in this run has limited effect, which is understandable given that it mainly acts on SNM.

Medians and 68\% CI of marginalized posterior distributions of key NM quantities, including those plotted in Fig.~\ref{Fig:Hist_NM}, are provided in Table~\ref{tab:Posteriors} in Appendix~\ref{App:Data}.

\subsubsection{Posterior distributions of global parameters of NS}
\label{sssec:NSParam}

\begin{figure*}
	\centering
	\includegraphics[]{"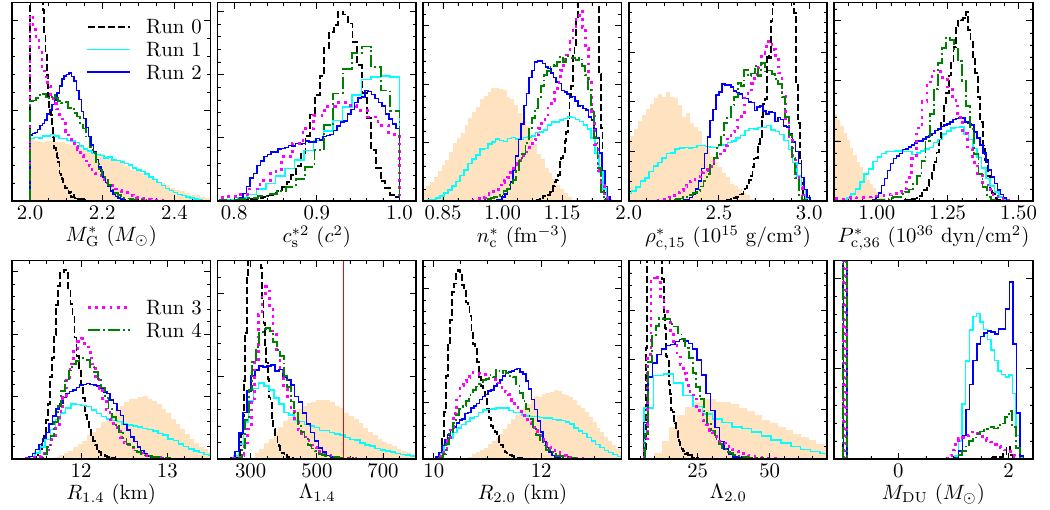"}
	\caption{Marginalized posteriors of selected properties of NSs. Considered are: maximum gravitational mass ($M_{\rm G}^*$); the central density corresponding to the most massive configuration ($n_{\mathrm c}^*$); speed of sound squared ($c_{\rm s}^{*2}$), energy density ($\rho_{\rm c}^*$) and pressure ($P_{\rm c}^*$) at $n_{\mathrm c}^*$; radii ($R_{1.4}$, $R_{2.0}$) and tidal deformabilities ($\Lambda_{1.4}$, $\Lambda_{2.0}$) of NSs with masses equal to $1.4~\Msun$ and $2.0~\Msun$; the lowest mass of a NS that accommodates for direct URCA ($M_{\rm DU}$). The vertical solid line on the $\Lambda_{1.4}$ plot illustrates the upper bound (at 90\% confidence) of the $190^{+390}_{-120}$ constraint from \citep{Abbott_PRL_121}; note that both the median value and the lower boundary sit outside the figure X-range. For $M_{\rm DU}$ the value ``$-1$'' corresponds to the models that do not allow this process to operate in stable NSs. For comparison, marginalized posteriors corresponding to run~4 of \citep{Beznogov_PRC_2023} are illustrated as well (orange shaded areas). Y-axis ranges have been chosen to increase readability.}
	\label{Fig:Hist_NS}
\end{figure*}

Equilibrium properties of non-rotating and spherically-symmetric NSs are inferred by solving, for each model of NS EOS, the Tolman–Oppenheimer–Volkoff equations. Tidal deformabilities are computed according to \cite{Hinderer_ApJ_2008,Hinderer_PRD_2010}. The lowest mass of a NS that accommodates for the direct URCA process ($M_{\rm DU}$) is the mass of the NS with $\rho_{\rm c}=\rho_{\rm DU}$, where the latter quantity is defined by $\Yp=Y_{\mathrm{p;\,DU}}$.

Posterior distributions of selected global properties of NSs are plotted on Fig.~\ref{Fig:Hist_NS}. The results of all five runs are illustrated. 

Runs~0 and 3 provide the two narrowest distributions of $M_{\rm G}^*$. In what regards run~0, the situation is expected as this is the case where by far the narrowest distributions of $P_\mathrm{NS}(n)$ is obtained, see Fig.~\ref{Fig:CD_NS}. The fact that for these runs the most probable value of $M_{\rm G}^*$ is $2~\Msun$, that is, the low limit imposed for the maximum mass, means that these runs gather the largest number of soft NS EOSs. As a consequence, NS matter can be strongly compressed, which explains on the one hand the large values obtained in the distributions of $n_{\mathrm c}^*$ and the related $\rho_{\rm c}^*$ and, on the other hand, the small values of $R_{1.4}$, $\Lambda_{1.4}$, $R_{2.0}$ and $\Lambda_{2.0}$. The narrowest (broadest) posteriors of $R_{1.4}$, $\Lambda_{1.4}$, $R_{2.0}$ and $\Lambda_{2.0}$ correspond to run~0 (1). Runs~0 and 2 -- 4 provide for $\Lambda_{1.4}$ relatively narrow posteriors, localized between the median value and the upper bound of the constraint $190^{+390}_{-120}$ (at 90\% confidence, \citealt{Abbott_PRL_121}) extracted from GW170817. The posterior of run~1 has a long tail whose extremity exceeds by 40\% the above mentioned upper limit from GW170817.

The widest distribution of $M_{\rm G}^*$ corresponds to run~1, in agreement with expectations based on the wide distribution of $P_\mathrm{NS}(n)$ in Fig.~\ref{Fig:CD_NS}. As a results, wide distributions characterize all other parameters plotted in Fig.~\ref{Fig:Hist_NS}, including $M_{\rm DU}$. The large number of models that allow for massive NSs explains why this is the only run where the $c_{\rm s}^{*2}$ distribution is peaked at 1, the largest accepted value.

We have seen in Fig.~\ref{Fig:CD_NS} that an extra condition on $\mN$ at $\ns$ or on $\epnm$ at a low density leads to NS EOSs stiffer that those of run~0. This explains why, for runs~2 and 4, the probability to have NS with $M_{\rm G}^*$ close to $2~\Msun$ is strongly reduced and the $M_{\rm G}^*$-distribution is peaked at larger values. The EOS stiffening gets translated into lower values of $n_{\rm c}^*$ and $\rho_{\rm c}^*$ and higher values of $R_{2.0}$ and $\Lambda_{2.0}$.

Posterior distributions of $M_{\rm DU}$ are particularly interesting in that in all runs there are models which allow for direct URCA as well as models where direct URCA is not allowed to operate in stable NSs. By far the smallest (largest) number of models which allow for direct URCA corresponds to run~0 (runs~1 and 2). This situation could be anticipated based on the values of $\asym{E}(n)$ in Fig.~\ref{Fig:CD_NM} even if in that figure the largest considered density is only a fraction of the largest value of $n_{\mathrm c}^*$ in Fig.~\ref{Fig:Hist_NS}. Out of all the models in run~1 that allow for direct URCA, 3.3\% accommodate it in NSs with masses smaller than $1.2~\Msun$. For run~2 the corresponding figure is 1.2\%. The variety of distributions of $M_{\rm DU}$ is an extra proof that each and every constraint acts on the density dependence of the symmetry energy and, thus, on NS particle composition.

Medians and 68\% CI of marginalized posterior distributions of key NS parameters, including some of those plotted in Fig.~\ref{Fig:Hist_NS}, are provided in Table~\ref{tab:Posteriors} in Appendix~\ref{App:Data}.

For comparison, also shown in Fig.~\ref{Fig:Hist_NS} are marginalized posteriors corresponding to run~4 of \citep{Beznogov_PRC_2023}, which is roughly equivalent with run~1 here. No posterior distribution is shown for $c_{\rm s}^{*2}$ as, with a maximum value of 0.77, the models in run~4 span a domain of values situated outside the X-axis range of our figure. No posterior distribution is shown for $M_{\mathrm{DU}}$, the reason being that direct URCA is not allowed in any of the models in run~4. All CDF distributions are shifted and of different shape with respect to those derived in this paper. The lowest (largest) values of $c_{\rm s}^{*2}$, $n_{\mathrm c}^*$, $\rho_{\rm c}^*$ and $P_{\rm c}^*$ ($R_{1.4}$, $\Lambda_{1.4}$, $R_{2.0}$ and $\Lambda_{2.0}$) in CDF indicate that these EOS models are considerably stiffer than those built within the non-relativistic theory of nuclear matter. Although not really visible in the figure,  the $M_{\rm G}^*$-distribution is noticeably wider for the CDF model than for any of the models in this work. Along with discrepancies in what regards the threshold for direct URCA, this corroborates the idea that the structure of the density functional prevails over constraints posed around $\sat{n}$.

\subsection{Correlations between parameters of NM}
\label{ssec:CorrelNM}

\begin{figure*}
	\centering
	\includegraphics[]{"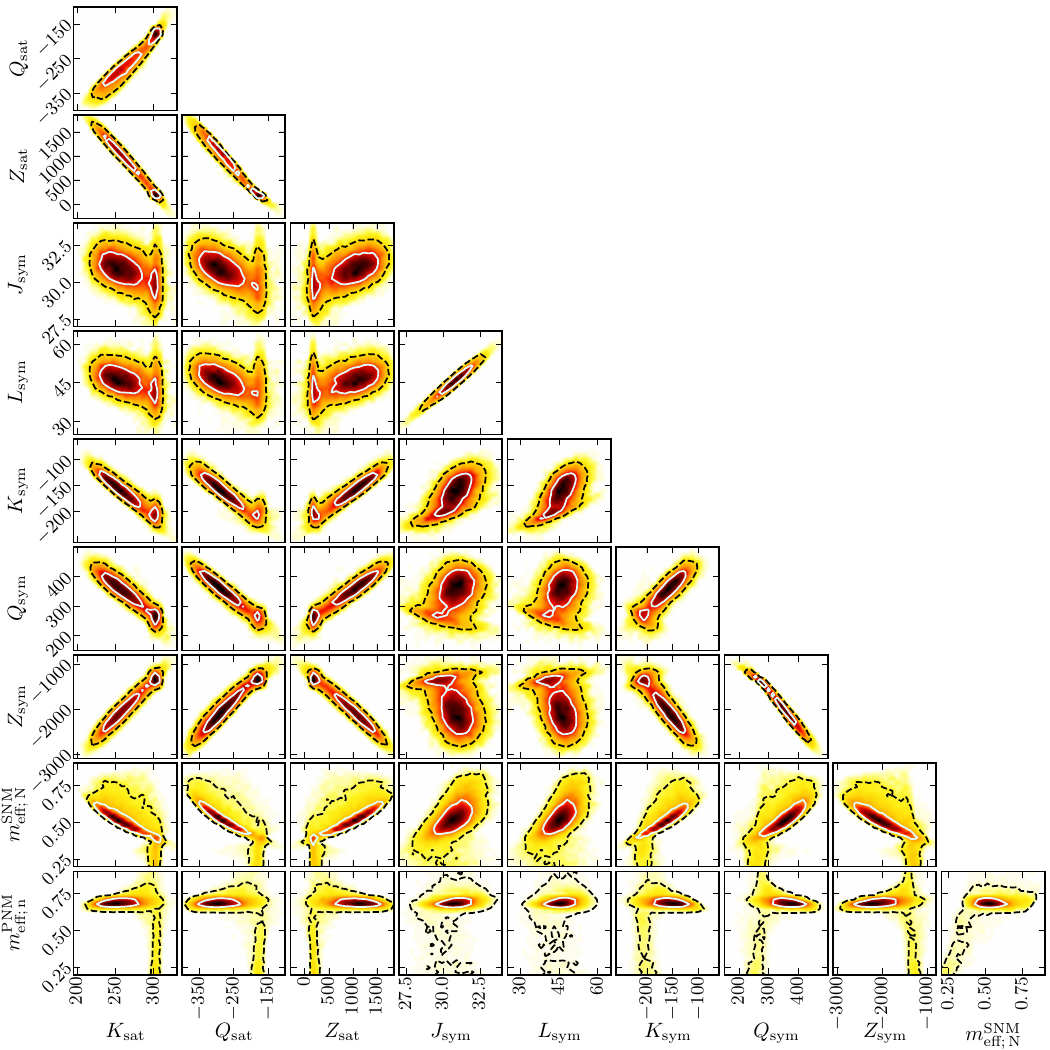"}
	\caption{Two-dimensional (2D) marginalized posteriors of some of the NM parameters. $\sat{X}$ and $\sym{X}$ parameters are expressed in MeV; $\mN$ and $\mn$ are expressed as fractions of the nucleon and neutron masses, respectively. The color map indicates the probability density. The light solid and the black dashed contours show 50\% and 90\% CR, respectively. Results correspond to run~0 in Table~\ref{tab:runs}.
	}
	\label{Fig:correl_NM_run0}
\end{figure*}

\begin{figure*}
	\centering
	\includegraphics[]{"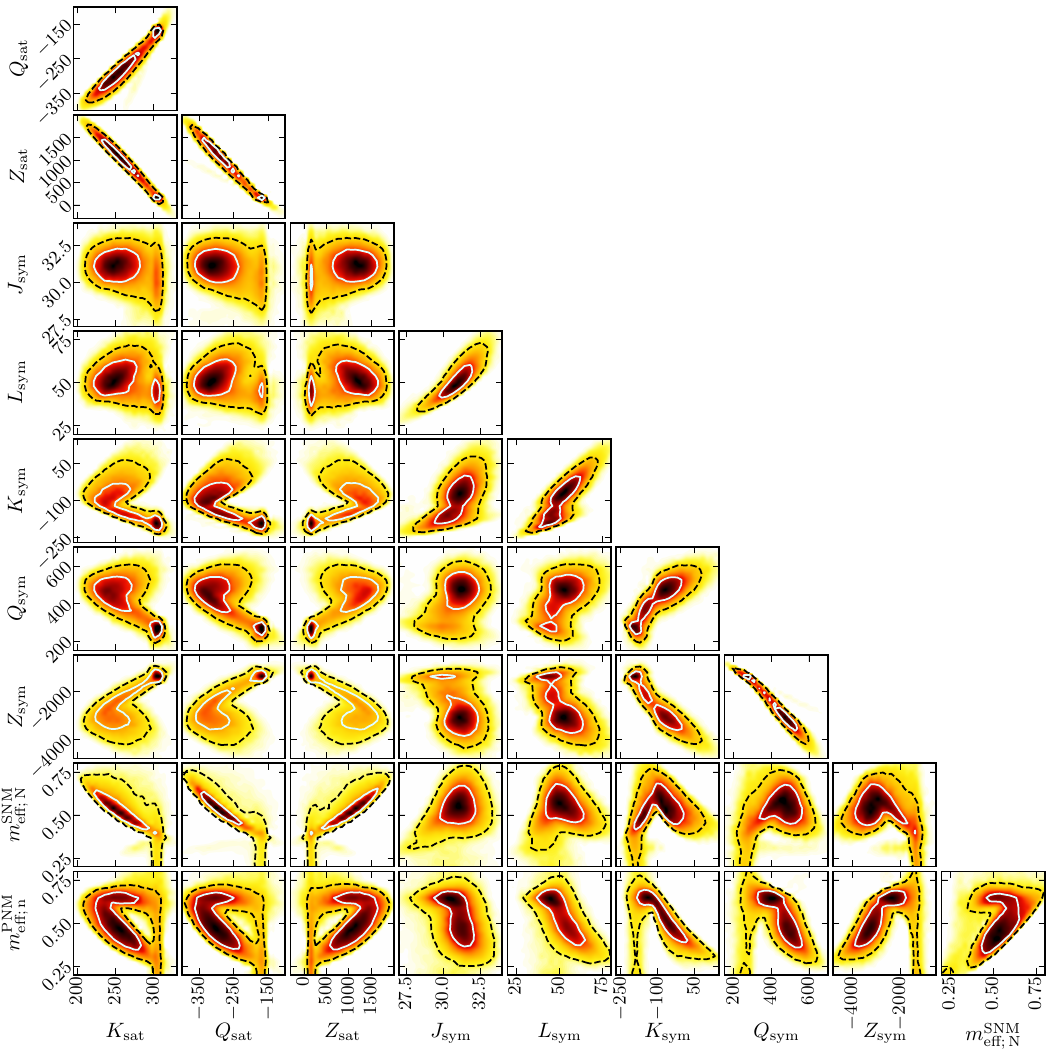"}
	\caption{The same as in Fig.~\ref{Fig:correl_NM_run0} but for run~1 in Table~\ref{tab:runs}.}
	\label{Fig:correl_NM_run1}
\end{figure*}

Correlations among parameters that describe the isoscalar and isovector behavior of the NM, $\mN$ and $\mn$ are investigated in Figs.~\ref{Fig:correl_NM_run0} and \ref{Fig:correl_NM_run1}, which correspond to run~0 and run~1 in Table~\ref{tab:runs}, respectively. Correlations with $\sat{n}$ and $\sat{E}$ are not included as the tight constraints imposed on these quantities do not allow for any correlation to manifest. 

It comes out that some of the correlations manifest themselves irrespective of whether correlations among $\left(E/A\right)_i$ with $i=2,3,4$ are accounted for, while other are sensitive to how the constraints from PNM are implemented. 

Examples for the first situation are provided by the positive (negative) correlation between $\sat{K}$ and $\sat{Q}$ ($\sat{K}$ and $\sat{Z}$); the negative correlation between $\sat{Q}$ and $\sat{Z}$; the negative correlation between $\sym{Q}$ and $\sym{Z}$. As mentioned in Sec.~\ref{sssec:NMParam}, these correlations manifest also in other models, which suggests that they might be universal. A positive correlation exists also between $\sym{J}$ and $\sym{L}$, and it becomes stronger when correlations among $\left(E/A\right)_i$ are included. We note that a $\sym{J}-\sym{L}$ correlation shows up also in \citep{Beznogov_PRC_2023}, while it does not in \citep{Malik_ApJ_2022,Malik_PRD_2023}; this situation is surprising the more considering that the model in \citep{Beznogov_PRC_2023} is a version of the model previously proposed in \citep{Malik_ApJ_2022}. We also note that, modulo the long tail at low values of $\mN$, both runs~0 and 1 show negative correlations between $\mN$ and $\sat{K}$. This result is easy to understand considering that $\mN$ is negatively correlated with $\eff{C}$ and $\eff{C}$ is positively correlated with $\sat{K}$, see Eq.~\eqref{eq:Ksat}. Runs~0 and 1 also show negative (positive) correlations between $\mN$ on the one hand and $\sat{Q}$ ($\sat{Z}$) on the other hand. The latter correlations are straightforward to explain based on the other correlations discussed above.

A spectacular outcome corresponds to correlations between the isoscalar and isovector channels $\sat{X}-\sym{Y}$, with $X,Y=K, Q, Z$ that exist for run~0 and do not exist for run~1. Notice that no strong cross-channel correlations have been identified by \cite{Malik_ApJ_2022,Beznogov_PRC_2023}, while correlations between $\sat{Q}$ and $\sym{Q}$ appear to be favored in CDF models with strong nonlinear vector $\omega$ field contribution \citep{Malik_PRD_2023}. A possible explanation for the absence of cross-channel correlations in the literature is that correlations among the values a function, which is used as a constraint, takes at different densities, are disregarded. If true, the situation might change if these correlations among the constraints are accounted for. Following the same reasoning, one might speculate that, by accounting for correlations among $\left(E/A\right)_i$, a $\sym{J}-\sym{L}$ correlation may appear in the CDF models of \cite{Malik_ApJ_2022,Malik_PRD_2023} and the correlation of \cite{Beznogov_PRC_2023} may become stronger. Alteration of correlation patterns upon inclusion of correlations among $\left(E/A\right)_i$ will be addressed in detail elsewhere. Until then, it is certain that a strong model dependence comes from the coupling between isovector and isoscalar channels. Experimental signatures of this ``connection'' would probably boost our understanding of the density dependence of the nuclear symmetry energy. Other examples of correlations that depend on how the constraints from $\left(E/A\right)_i$ are implemented are provided by the positive (negative) correlations between $\mN$ and $\sym{J}$, $\sym{L}$, $\sym{K}$, $\sym{Q}$ ($\sym{Z}$), which exist for run~0 and do not exist for run~1; the correlations $\mN-\sym{J}$ and $\mN-\sym{L}$ are weak and related one to another by the strong correlation $\sym{L}-\sym{J}$; the correlations $\mN-\sym{X}$ with $X=K,Q,Z$ can be viewed as the consequence of the correlations $\mN-\sat{K}$ and correlations $\sat{K}-\sym{X}$.

Results of this section can be cross-checked with those of \cite{Beznogov_PRC_2023}, where a study similar to the one performed here was done using EOSs built within the CDF theory with simplified density-dependent couplings. Note nevertheless that, for all quantities that have been constrained, the values of the constraints differ slightly from those used in this work. Moreover, there is only one case that, modulo these slight differences, is common for our paper \citep{Beznogov_PRC_2023} and the present work. That is run~4 of \cite{Beznogov_PRC_2023}, which is equivalent to run~1 in this paper. The fiducial run of \cite{Beznogov_PRC_2023}, to which Figs.~2 and 3 of that paper correspond, additionally accounts for constraints on the pressure in PNM at three values of the density. As such it can be considered somehow intermediate between runs~1 and 0 here. For a complementary discussion about correlation dependence on theoretical framework; effective interaction; size of domains; progressive incorporation of constraints, the reader is referred to Sec.~III of \cite{Beznogov_PRC_2023} as well as to \cite{Fattoyev_PRC_2012, Papakonstantinou_PRC_2023, Malik_PRD_2023}.

%


\subsection{Correlations between parameters of NM and properties of NSs}
\label{ssec:CorrelNM-NS}

\begin{figure*}
	\centering
	\includegraphics[]{"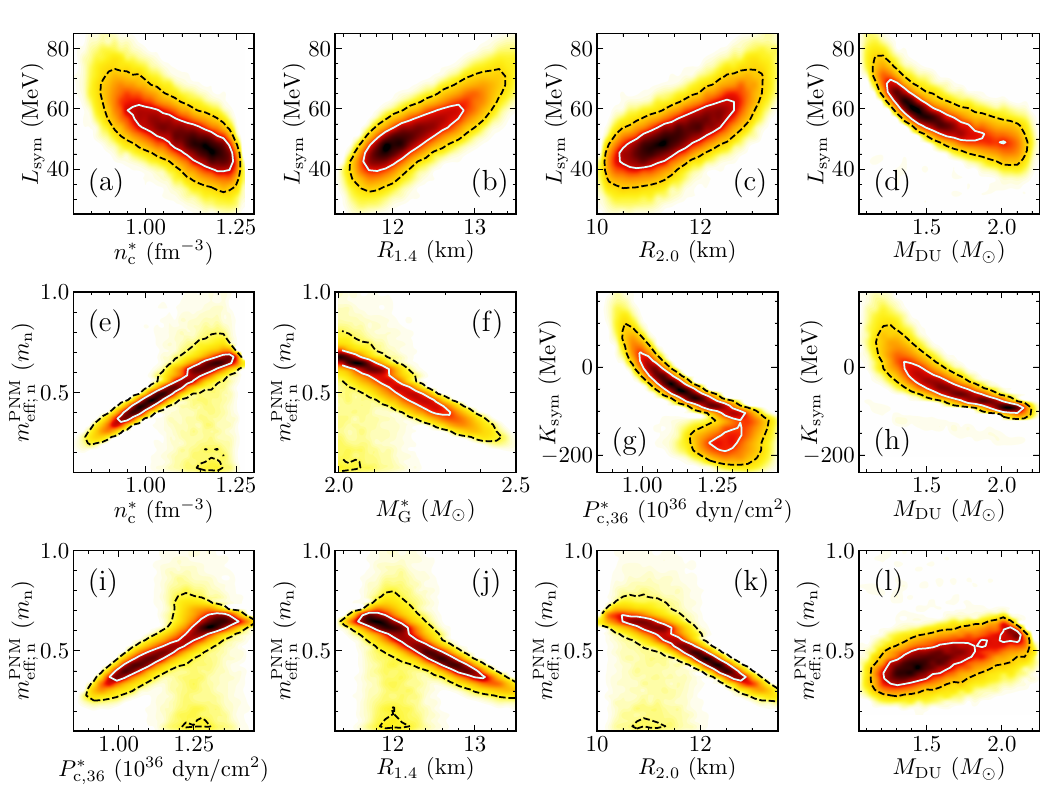"}
	\caption{2D marginalized posterior distributions for selected NM and NS parameters for which correlations are obtained. The color map indicates the probability density. The light cyan solid and the black dashed contours demonstrate 50\% and 90\% CR, respectively. Results correspond to run~1 in Table~\ref{tab:runs}.
	}
	\label{Fig:Correl_NS-NM_run1}
\end{figure*}

Correlations between parameters of NM and properties of NSs are considered in Fig.~\ref{Fig:Correl_NS-NM_run1} for run~1.

It comes out that $R_{1.4}$ and, to a lesser extent, $R_{2.0}$ are positively correlated with $\sym{L}$. Correlations occur also between $R_{1.4}$ and $R_{2.0}$ and $\sym{K}$, but are not shown on this figure. For $\sym{K} \gtrsim -100~\mathrm{MeV}$, $\sym{K}$ is negatively correlated with $P_{\mathrm c}^*$. These results agree with the conclusions of \cite{Fortin_PRC_2016, Papakonstantinou_PRC_2023}, where both phenomenological non-relativistic and relativistic mean field models have been considered, and are expected to manifest also in other models. The rationale is that large values of $\sym{L}$ and $\sym{K}$ engender stiff NS EOSs that prevent matter from being compressed too much. Limited compression translates into low values of $n_{\mathrm c}^*$ and $P_{\mathrm c}^*$ and large values of $R_{1.4}$ and $R_{2.0}$, in agreement with panels (a),  (b), (c) and (g), respectively.

$M_\mathrm{DU}$ appears to be negatively correlated with $\sym{L}$ [panel (d)] and $\sym{K}$ [panel (h)]. The first of these correlations was previously commented by \cite{Fortin_PRC_2016}. Both correlations are easy to understand and, as before, are expected to be present also in other models. The explanation is that large values of $\sym{L}$ and $\sym{K}$ lead to large values of $\sym{E}\left(n\right)$ that prevent $\beta$-equilibrated matter from getting depleted from its protons. The fact that none of these correlations are strong is attributable to higher order parameters that are negatively correlated with $\sym{L}$ and $\sym{K}$, as it is the case of $\sym{Z}$ (see Fig.~\ref{Fig:correl_NM_run1}). The strengths of these correlations nevertheless depend on the details of the model.

According to Fig.~\ref{Fig:Correl_NS-NM_run1} $\mn$ at $\ns$ appears to have an impact also on $n_{\mathrm c}^*$ [panel (e)], $M_{\mathrm G}^*$ [panel (f))], $P_{\mathrm c}^*$ [panel (i)], $R_{1.4}$ [panel (j)], $R_{2.0}$ [panel (k)] and  $M_\mathrm{DU}$ [panel (l)].  To our knowledge none of these correlations were noticed before. 

We underscore that, when correlations among $E/A$ in PNM are accounted for, only the $\sym{K}-M_{\mathrm DU}$ correlation survives. A plausible explanation for this situation is that the domains where the various quantities take values shrink too much too allow correlations to manifest, see Figs.~\ref{Fig:Hist_NM} and \ref{Fig:Hist_NS}.


\section{Summary}
\label{sec:Summary}

The non-relativistic mean field theory of nuclear matter and the standard Skyrme parametrization of the nucleonic effective interaction have been employed to generate, within a Bayesian framework, models for NM and NS EOS. Several types and different combinations of constraints have been used. In the first place constraints were posed on the four best known NM properties, i.e., saturation density of symmetric matter; energy per nucleon and compressibility of symmetric saturated matter; symmetry energy at saturation. The considered values were those obtained by \cite{Margueron_PRC_2018a}, based on a compilation of 35 standard Skyrme interactions frequently used in the literature. Then, the behavior of PNM was controlled by the values that $\epnm$ takes in a minimum of three and a maximum of four points with densities in the range $0.04~\mathrm{fm}^{-3} \leq n \leq \ns$. For $\epnm$ values obtained within $\chi$EFT with NN at N$^3$LO and 3N at N$^2$LO were used \citep{Drischler_PRC_2021}. Finally, models of NSs were required to comply with the $2~\Msun$ condition on the maximum mass and be causal at the density corresponding to the central density of the maximum mass configuration. The effectiveness of constraints from PNM was tested by accounting for or disregarding the correlations among the values that $\epnm$ takes at different densities; implementing constraints at three or four densities. Two sets of models also implemented constraints on the values of the nucleon (neutron) Landau effective masses in SNM (PNM) at $\ns$; similarly to $\epnm$, the values obtained by \cite{Drischler_PRC_2021} were used.

The wide domain of values associated with the constraint on $\sat{K}$ together with the limited effect the constraints on $\epnm$ have on SNM get reflected in a significant uncertainty in what regards the high density behavior of $E/A \left( n \right)$ in SNM and an expected similitude among the domains of values that the models in each run span for a fixed density. At variance with this, the behavior of PNM differs much from one run to the other. The smallest uncertainties in the behavior of PNM as a function of density correspond to the situations where correlations among $\epnm$ at various densities have been accounted for. The largest uncertainties correspond to the opposite case of minimal number of constraints on $\epnm$. Largely different behaviors of PNM get translated into largely different behaviors of the density dependence of $\asym{E}$ and NS EOS. We note, in particular, that the softest NS EOS, though still compatible, by construction, with the $2~\Msun$ limit, corresponds to the case where correlations among $\left(E/A\right)_i$ in PNM were implemented; the largest values of the maximum mass, of the order of $2.5~\Msun$, were obtained for models in the least constrained run. Large uncertainties in $\asym{E}$, due to large uncertainties in both SNM and PNM, make that all runs contain models where the (a)symmetry energy increases steeply with the density as well as models where this quantity vanishes at densities a few times larger than the saturation density. Models in the first category allow the direct URCA to operate in stars as light as $\approx 1~\Msun$, while models in the second category do not allow at all for this fast neutrino emission process.  

Conditional probability density distributions $P(R|M)$ show that, for each of our runs, a significant number of models comply with joint mass and radius posterior probability density contours corresponding to PSR J0030+0451 \citep{Miller_2019} and PSR J0740+6620 \citep{Miller_may2021}. In contrast, a very small number of models appear to comply with joint mass and radius measurements of the NS in the SNR HESS J1731–34 \citep{Doroshenko_Nature_2022}; moreover, none of them agrees with constraints from the other two observations above. At the very least, this tension can be attributed to a tension between HESS J1731–34 data and data on $\epnm$ from $\chi$EFT \citep{Drischler_PRC_2021} or, rather, a limited flexibility of the energy density functional.

In what regards the combined tidal deformability as a function of mass ratio it turns out that the models in the most constrained run provide values by $\approx 30\%$ higher than the median value extracted from GW170817~\citep{Abbott_PRX_2019}. At variance with this, models in the least constrained run provide values that span a wide domain; their median value is by $60\%-70\%$ larger than the median value in \citep{Abbott_PRX_2019}; the 90\% upper bound of the CR of models in this run is close to the 90\% symmetric upper bound in \citep{Abbott_PRX_2019}. Overall, our results are fully compatible with GW170817 data.  

Marginalized posterior distributions of all NEPs except $\sat{n}$ and $\sat{E}$; $\epnm$ over $0.04~\mathrm{fm}^{-3} \leq n \leq \ns$; $\mn$; $\mN$ differ considerably from one run to the others, which means that every constraint is effectual. Target distributions of $\sat{K}$ and $\epnm$ at $\ns$ are never met, which, again, might signal either an incompatibility among constraints or the limited flexibility of the energy density functional. Target distributions of $\mn$ and $\mN$ are met only when constraints on these quantities are imposed; in all other cases their marginalized posteriors are broad and shifted towards lower values. 

Marginalized posterior distributions of various NS properties also differ from one run to the others; the largest (narrowest) distributions are obtained for the least (most) constrained runs (runs~1 and 0, respectively).
When compared with equivalent distributions obtained within Bayesian inferences where EOS models are built based on the CDF~\citep{Malik_ApJ_2022,Beznogov_PRC_2023}, considerable discrepancies arise. First, Skyrme-like interactions lead to NS EOS that are softer than their CDF counterparts, which gets reflected in NS with increased compactness and smaller radii and tidal deformabilities. The largest values of matter density met in the NS's inner cores together with the steep increase of $\sym{E}(n)$ make that between 11\% and 58\% of models belonging to runs 1 to 4 allow the direct URCA to operate in stable stars. For run~0, which produces the largest $n_{\mathrm c}^{*}$ but has the lowest symmetry energy at high densities, direct URCA operates only in 1\% of models and in NS with masses close to the maximum mass. This situation is at variance with the one seen in CDF with density dependent couplings, where this fast cooling process is completely forbidden. The obvious conclusion is that, in what regards the high density behavior, the role played by the density functional dominates the one played by constraints imposed around $\sat{n}$.

Correlations among NS EOS stiffness; maximum mass as well as corresponding values of the central (energy) density and pressure; radii and tidal deformabilities of NS with $M/\Msun=1.4,~2$ are in agreement with previous findings in the literature. In accord with some models in the literature, our models show negative correlations between $M_\mathrm{DU}$ on one hand and $\sym{L}$ and $\sym{K}$ on the other hand. We also find positive (negative) correlations between $\mn$ and $n_{\mathrm c}^*$, $P_{\mathrm c}^*$, $M_\mathrm{DU}$ ($M_\mathrm{G}$, $R_{1.4}$, $R_{2.0}$). None of those have been noted before and are worth more investigation.
 
The run that accounts for correlations among $\left( E/A \right)_i$ in PNM is unique in manifesting strong cross channel correlations between the parameters that describe the isoscalar and isovector behavior of the NM EOS; correlations between $\mN$ and $\sym{X}$; a strong correlation $\sym{J}-\sym{L}$. As discussed at length in the literature, correlations between NEPs or between NEPs on the one hand and properties of NS on the other hand depend on the employed theoretical framework; model of effective interaction; domains of values that input and physical parameters are allowed to span; constraints. In order for a correlation to be considered physical it has to manifest in every model where each of the quantities it involves spans a wide enough domain of values. In view of these, it appears that the positive correlations between $\sat{K}-\sat{Q}$, $\sym{K}-\sym{Q}$ and the negative correlations between $\sat{K}-\sat{Z}$, $\sat{Q}-\sat{Z}$, $\sym{K}-\sym{Z}$, $\sym{Q}-\sym{Z}$ are physical. The situation of other correlations, e.g., $\sym{J}-\sym{L}$ is more ambiguous.

\section{Conclusions}
\label{sec:Concl}

In this work we have performed a \emph{full} Bayesian investigation of the dense matter EOS built within the non-relativistic approach of nuclear matter with standard Skyrme interactions. A strategy and constraints similar to those previously used by \cite{Malik_ApJ_2022,Beznogov_PRC_2023,Malik_PRD_2023} have been adopted. The ensemble of these models adds to a more thorough investigation of the still unknown NM and NS EOSs. Confrontation of their results contributes to tracking the model dependence of their conclusions, the physical character of some correlations between potentially measurable global parameters of NS or with parameters of NM, as well as the efficiency of every constraint. 

The consequences of accounting for correlations among the values that $E/A$ in PNM takes at different values of density~\citep{Drischler_PRC_2021} have been carried out here for the first time. In addition to being very instrumental in reducing the uncertainty range in the isovector channel, these correlations have been shown to couple the isovector and isoscalar channels.  It remains to test whether the last feature is a peculiarity of Skyme or a more general outcome.

Constraints imposed on nucleon effective masses at $\sat{n}$ also affect the density behavior of the EOS, though to a limited extent. Much stronger effects are expected if constraints are imposed on the values that $m_{\mathrm{eff};i}$ take over a finite density domain. This aspect is under investigation and will be addressed elsewhere.

\begin{acknowledgments}
	The authors thank Sergey Koposov and Johannes Buchner for useful discussion regarding the implementation specifics of Nested Sampling in \textsc{dynesty} and \textsc{ultranest} Python packages. The authors also appreciate the data and useful comments provided by Victor Doroshenko regarding the NS in SNR HESS J1731–34.
	
	Support from a grant of the Ministry of Research, Innovation and Digitization, CNCS/CCCDI – UEFISCDI, Project No. PN-III-P4-ID-PCE-2020-0293 is also acknowledged. Partial support from PN 23 21 01 02 is acknowledged as well.
	
	M.V.B. and A.R.R. contributed equally to this work.
\end{acknowledgments}
\software{
	\begin{itemize}[nosep]
		\item Private codes written by authors
		\item  \textsc{Python} [\url{https://www.python.org}] with packages:
		\begin{itemize}[nosep]
			\item \textsc{matplotlib} [\cite{Matplotlib}, \url{https://matplotlib.org}] 
			\item \textsc{numpy} [\cite{NumPy}, \url{https://numpy.org}]
			\item \textsc{scipy} [\cite{SciPy}, \url{https://scipy.org}]
			\item \textsc{emcee} [\cite{emcee,emcee3}, \url{https://github.com/dfm/emcee}]
			\item \textsc{ultranest} [\cite{ultranest}, \url{https://github.com/JohannesBuchner/UltraNest}]
		\end{itemize}
	\end{itemize}
}
\appendix
\section{Nuclear Empirical Parameters}
\label{App:NEPs}

We provide here analytic expressions for low order NEPs in terms of $C_0$, $D_0$, $C_3$, $D_3$, $\eff{C}$, $\eff{D}$ and $\sigma$.
For equivalent expressions in terms of the most commonly used $t_0$, $x_0$, $t_1$, $x_1$, $t_2$, $x_2$, $t_3$, $x_3$ and $\sigma$ coefficients, see \cite{Dutra_PRC_2012}.

The saturation energy of matter with isospin asymmetry $\delta$ writes: 
\onecolumngrid
\begin{align}
	\begin{split}
		\sat{E} \left(\delta\right) &= \frac{3 \hbar^2}{10 m} \left(\frac{3 \pi^2}{2} \right)^{2/3} G_{5/3}(\delta) \left(\sat{n}^{\delta}\right)^{2/3} + \left( C_0+D_0 \delta^2\right)\sat{n}^{\delta} + \left( C_3+D_3 \delta^2\right) \left(\sat{n}^{\delta}\right)^{\sigma+1}  \\
		&+\frac{3}{10} \left(\frac{3 \pi^2}{2} \right)^{2/3} \left[ \eff{C} \left( \left( 1 + \delta \right)^{5/3}+\left(1-\delta\right)^{5/3} \right) +\eff{D} \delta \left( \left( 1 + \delta \right)^{5/3}-\left(1-\delta\right)^{5/3} \right) \right] \left(\sat{n}^{\delta}\right)^{5/3},
	\end{split}
    \label{eq:Esat} 
\end{align}
The expressions for incompressibility $\sat{K}\left(\delta\right)=\sat{X}^{\delta;\,2}$, skewness $\sat{Q}\left(\delta\right)=\sat{X}^{\delta;\,3}$ and kurtosis $\sat{Z}\left(\delta\right)=\sat{X}^{\delta;\,4}$ of NM with isospin asymmetry $\delta$ are:
\begin{align}
	\begin{split}
		\sat{K}\left(\delta\right)&= \left. \left[-18 \frac{P}{n}+9\frac{\partial P}{\partial n}\right] \right|_{n=\sat{n}^{\delta}} \\
		&= \frac{3 \hbar^2}{m} \left( \frac{3 \pi^2}{2}\right)^{2/3} G_{5/3}(\delta) \left(\sat{n}^{\delta}\right)^{ 2/3}	+18 \left( C_0+D_0 \delta^2\right)\sat{n}^{\delta}	+ 9 \left( \sigma+1\right) \left( \sigma+2\right) \left( C_3+D_3 \delta^2\right) \left(\sat{n}^{\delta}\right)^{\sigma+1}  \\
		&+12 \left( \frac{3 \pi^2}{2}\right)^{2/3} \left[ \eff{C} \left( \left(1+\delta\right)^{5/3}+\left(1-\delta\right)^{5/3} \right) + \eff{D} \delta \left( \left(1+\delta\right)^{5/3}-\left(1-\delta\right)^{5/3} \right) \right] \left(\sat{n}^{\delta}\right)^{5/3},
	\end{split}
    \label{eq:Ksat} \\
	\begin{split}
		\sat{Q}\left(\delta\right) & = 3 \sat{n}^{\delta} \left. \frac{\partial K \left( n \right)}{\partial n} \right|_{n=\sat{n}^{\delta}} - 3 \cdot 2 \sat{K}\left(\delta\right)  \\
		&=\frac{12 \hbar^2}{5 m} \left( \frac{3 \pi^2}{2}\right)^{2/3} G_{5/3}(\delta) \left(\sat{n}^{\delta}\right)^{2/3} +27 \left( \sigma+1\right) \sigma \left( \sigma-1\right) \left( C_3+D_3 \delta^2\right) \left(\sat{n}^{\delta}\right)^{\sigma+1} \\
		&-3 \left( \frac{3 \pi^2}{2}\right)^{2/3} \left[ \eff{C} \left( \left( 1+\delta\right)^{5/3} + \left( 1-\delta\right)^{5/3} \right) + \eff{D} \delta \left( \left( 1+\delta\right)^{5/3} - \left( 1-\delta\right)^{5/3} \right) \right] \left(\sat{n}^{\delta}\right)^{5/3},
	\end{split}
    \label{eq:Qsat} \\
	\begin{split}
		\sat{Z}\left(\delta\right) & =  3 \sat{n}^{\delta} \left.\frac{\partial Q \left( n \right)}{\partial n} \right|_{n=\sat{n}^{\delta}} - 3 \cdot 3 \sat{Q}\left(\delta\right)  \\
		& = -7 \frac{12 \hbar^2}{5 m} \left( \frac{3 \pi^2}{2}\right)^{2/3} G_{5/3}(\delta) \left(\sat{n}^{\delta}\right)^{2/3} + 3 \cdot 27 \left( \sigma+1\right)\sigma \left( \sigma-1\right) \left( \sigma - 2 \right) \left( C_3+D_3 \delta^2\right) \left(\sat{n}^{\delta}\right)^{\sigma+1}  \\
		& + 12 \left( \frac{3 \pi^2}{2}\right)^{2/3} \left[ \eff{C} \left( \left( 1+\delta\right)^{5/3} + \left( 1-\delta\right)^{5/3} \right) +\eff{D} \delta \left( \left( 1+\delta\right)^{5/3} - \left( 1-\delta\right)^{5/3} \right) \right] \left(\sat{n}^{\delta}\right)^{5/3}.
	\end{split}
   \label{eq:Zsat}
\end{align}
In the above expressions we made use of $\tau_i(T=0)=\pi^{4/3} \left(3 n_i\right)^{5/3}/5$ and introduced $G_\alpha(\delta)=\left[ \left(1-\delta \right)^\alpha+\left(1+\delta \right)^\alpha\right]/2$. 

The symmetry energy can be computed as
\begin{align}
	\begin{split}
		E_{\mathrm{sym};\,2}(n)  = \frac12 \left. \frac{ \partial^2 \left(e/n \right)}{\partial \delta^2}\right|_{n,\,\delta=0}
		 = \frac{\hbar^2}{6 m} \left( \frac{3 \pi^2}{2}\right)^{2/3} n^{2/3} + D_0 n+D_3 n^{\sigma+1} +\left( \frac{\eff{C}}{3}+\eff{D}\right)  \left( \frac{3 \pi^2}{2}\right)^{2/3} n^{5/3},
	\end{split}
	\label{eq:Esym2}
\end{align}
while its slope $\sym{L}=3 \sat{n}^0 \left. \left( \partial E_{\mathrm{sym};\,2}(n)/\partial n \right) \right|_{n=\sat{n}^0}$,
curvature $\sym{K}=3^2 \left(\sat{n}^0\right)^2 \left. \left( \partial^2 E_{\mathrm{sym};\,2}(n) / \partial n^2 \right) \right|_{n=\sat{n}^0}$,
skewness $\sym{Q}=3^3 \left(\sat{n}^0\right)^3 \left. \left(\partial^3 E_{\mathrm{sym};\,2}(n) /\partial n^3 \right) \right|_{n=\sat{n}^0} $
and kurtosis $\sym{Z}=3^4 \left(\sat{n}^0\right)^4 \left. \left(\partial^4 E_{\mathrm{sym};\,2}(n) /\partial n^4 \right) \right|_{n=\sat{n}^0} $ write
\begin{align}
	&\sym{L} = \frac{\hbar^2}{3 m} \left( \frac{3 \pi^2}{2}\right)^{2/3} \sat{n}^{2/3} +3 D_0 \sat{n} +3 D_3 \left(\sigma+1 \right) \sat{n}^{\sigma+1} + 5 \left( \frac{\eff{C}}{3}+\eff{D}\right) \left( \frac{3 \pi^2}{2}\right)^{2/3} \sat{n}^{5/3},
	\label{eq:Lsym} \\
	&\sym{K} = -\frac{\hbar^2}{3 m} \left( \frac{3 \pi^2}{2}\right)^{2/3} \sat{n}^{2/3} +3^2 D_3 \sigma \left( \sigma+1\right) \sat{n}^{\sigma+1} +10 \left( \frac{\eff{C}}{3}+\eff{D}\right) \left( \frac{3 \pi^2}{2}\right)^{2/3} \sat{n}^{5/3},
	\label{eq:Ksym} \\
	&\sym{Q} = \frac{4 \hbar^2}{3 m} \left( \frac{3 \pi^2}{2}\right)^{2/3} \sat{n}^{2/3} +3^3 D_3 \left( \sigma+1\right) \sigma \left( \sigma-1\right) \sat{n}^{\sigma+1} -10 \left( \frac{\eff{C}}{3}+\eff{D}\right) \left( \frac{3 \pi^2}{2}\right)^{2/3} \sat{n}^{5/3},
	\label{eq:Qsym} \\
	&\sym{Z} = -\frac{4 \cdot 7 \hbar^2}{3m} \left( \frac{3 \pi^2}{2}\right)^{2/3} \sat{n}^{2/3} +3^4 D_3 \left( \sigma+1\right) \sigma \left( \sigma-1\right) \left( \sigma-2\right) \sat{n}^{\sigma+1} +40  \left( \frac{\eff{C}}{3}+\eff{D}\right) \left( \frac{3 \pi^2}{2}\right)^{2/3} \sat{n}^{5/3}.
	\label{eq:Zsym}
\end{align}
\twocolumngrid
We note that $\sym{J}$, $\sym{K}$, $\sym{Q}$, $\sym{Z}$ depend explicitly only on $\sigma$, $D_0$, $D_3$, $\eff{D}$ and $\eff{C}$. Notice that in Eqs.~\eqref{eq:Lsym} -- \eqref{eq:Zsym} the ``0" superscript has been dropped off for convenience. The same convention is used throughout this paper.


\section{MCMC implementation details}
\label{App:MCMC}

Similarly to our paper \citep{Beznogov_PRC_2023}, we have chosen \textsc{emcee} (v.3.1.4) Python package \citep{emcee,emcee3}\,\footnote{The documentation is available at \url{https://emcee.readthedocs.io} and the GitHub page is \url{https://github.com/dfm/emcee}.} as an implementation of the affine invariant MCMC \citep{Goodman_2010} and as our main Bayesian inference tool. In all calculations we have employed KDE (kernel density estimate) steps and 1\,000 walkers. While KDE steps are slower than ``classical'' stretch steps, the decrease in the integrated autocorrelation time compensates the increase in the time needed to compute one step. 

Our initial tests indicated that the MCMC sampling was failing, the Markov chains were not converging. Further investigation demonstrated that statistical analysis in terms of effective interaction input parameters $C_0,D_0,C_3,D_3,\eff{C},\eff{D}$ and $\sigma$ was all but impossible. Random uniform sampling from the prior showed that $\approx 97.8\%$ of the models had saturation density of SNM far outside the target range. Thus, a re-parametrization was necessary. Luckily, there are simple analytical expressions that relate effective interaction parameters with NEPs, Eqs.~\eqref{eq:re-param}. Consequently, we have employed the ``mixed'' parametrization with 3 NEPs and 4 effective interaction parameters, as described in Sec.~\ref{ssec:Prior}. In principle, there was a possibility to use 4 NEPs and 3 effective interaction parameters parametrization by adding $\sat{K}$ to the list of NEPs and removing $D_3$ from the list of effective interaction parameters. This would have improved the convergence even more. However, involving  $\sat{K}$ leads to Eqs.~\eqref{eq:re-param} being singular not only at $\sigma = 0$, which we have easily circumvented by adjusting the prior range, but also at $\sigma = 2/3$. This was problematic to circumvent as it would have required to split the prior range of $\sigma$ into something like $[0.01, 2/3 - \epsilon] \cup [2/3 + \epsilon, 1.1]$. Taking into account that the posterior distributions of $\sigma$ for most of the runs show a peak near $\sigma = 2/3$ (Fig.~\ref{Fig:Hist_ModelParam}), we decided against  adding $\sat{K}$ to the input parameters despite better convergence. Notice also that any re-parametrization changes the prior distributions unless the new prior distributions specifically counteract this change. We have not introduced such corrections to the prior distributions and used uniform prior distributions. Instead, we have compared both parametrizations against each other on a test model, see details below.

Even after the re-parametrization, the efficiency of the MCMC sampling was too low to be practical (in the sense of computational time) due to very high integrated autocorrelation lengths ($\tau_\mathrm{orig} \approx 500 - 10\,000$ depending on the run) and very low acceptance fractions ($0.5\%-1.5\%$). Very long chains were required to overcome these two issues. A different strategy was needed. The structure of the imposed constraints allows for a ``natural'' separation between the ``probabilistic'' ones handled via Eqs.~\eqref{eq:Chi2-NoCorr} and \eqref{eq:Chi2-Corr} and the ``hard wall'' ones. The latter constraints can be viewed as post-filtering criteria for the posterior obtained by imposing only the former constraints. Moreover, one call of the log-likelihood requires $\approx 1$~ms, while the time needed to compute $M_\mathrm{G}^*$ once is $\approx 1$~s (on the same machine). Thus, we have split the calculation into two parts. First, we compute the posterior via MCMC without $M_\mathrm{G}^*$ and $c_\mathrm{s}$ constraints. While this is as inefficient as before, it is now $\approx 1000$ times faster as we do not compute $M_\mathrm{G}^*$. This makes the first part reasonable in terms of computation time. Then, in principle, we need to post-filter the results by accepting only the models that comply with constraints on $M_\mathrm{G}^*$ and $c_\mathrm{s}$. Yet, if we did just that, it would have taken the same (very long) amount of time as running MCMC directly without splitting the constraints. Instead, we have ``thinned'' the chain by taking only approximately one sample per the integrated autocorrelation length before post-filtering. We aimed for the integrated autocorrelation length of the ``thinned'' chain to be $\tau_\mathrm{th} = 1.0-1.2$. By doing so we have simultaneously reduced the computational time for post-filtering by a factor of $\approx \tau_\mathrm{orig}$ and ensured that the ``thinned'' chain contained only (almost) \emph{independent} samples. ``Thinning'' also alleviates the issue of low acceptance fraction of the MCMC sampling. The acceptance fraction of the post-filtering itself varied from $\approx 0.8\%$ to $\approx 50\%$ depending on the run. As a result of such two stage ``thinning'', we can be certain that the final posterior contains only independent samples. 

\renewcommand{\arraystretch}{1.10}
\setlength{\tabcolsep}{8.0pt}
\begin{table}
	\caption{The ranges of the priors for run~1b.}
	\centering
	\begin{tabular}{cccc}
		\toprule
		\toprule
		Parameter      & Units                          & Min.              & Max.             \\
		\midrule
		$C_0$          & MeV $\mathrm{fm^{3}}$          & $-2.0\times 10^3$ & $4.3\times 10^3$ \\
		$D_0$          & MeV $\mathrm{fm^{3}}$          & $-4.4\times 10^3$ & $6.0\times 10^3$ \\
		$C_3$          & MeV $\mathrm{fm^{3+3\sigma}}$  & $-6.6\times 10^3$ & $1.9\times 10^3$ \\
		$D_3$          & MeV $\mathrm{fm^{3+3\sigma}}$  & $-6.0\times 10^3$ & $3.0\times 10^3$ \\
		$\eff{C}$      & MeV $\mathrm{fm^{5}}$          & 0.0               & 900.0            \\
		$\eff{D}$      &  MeV $\mathrm{fm^5}$           & $-900$            & 900.0            \\
		$\sigma$       &  --                            & 0.1               & 1.1              \\
		\bottomrule
		\bottomrule
	\end{tabular}
	\label{tab:Prior-1b}
\end{table}
\setlength{\tabcolsep}{2.0pt}
\renewcommand{\arraystretch}{1.0}

%
\begin{figure*}
	\centering
	\includegraphics[]{"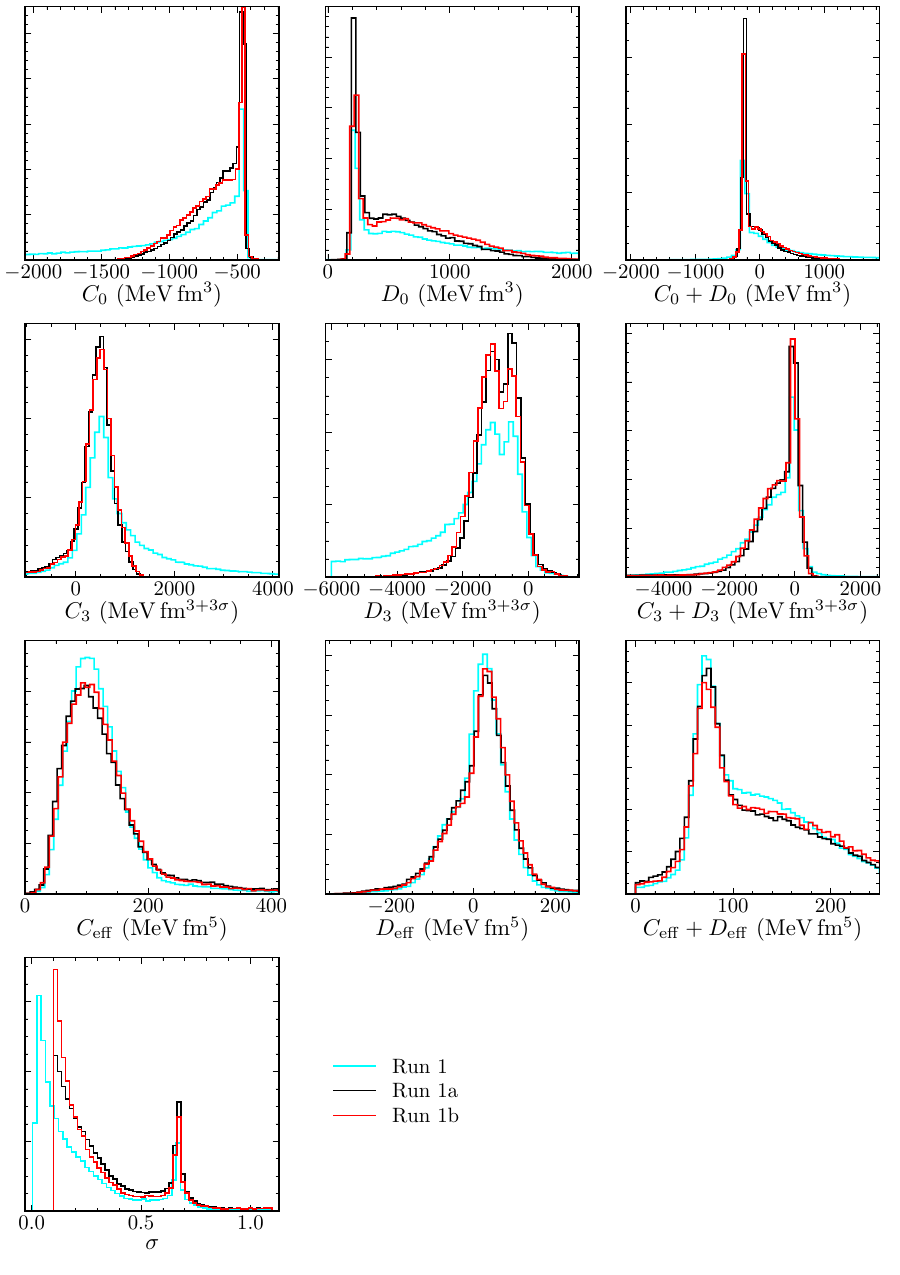"}
	\caption{Marginalized posterior distributions for the parameters of standard Skyrme effective interactions. Also shown are marginalized posteriors for $\left(C_0+D_0 \right)$, $\left(C_3+D_3 \right)$, $\left(\eff{C}+\eff{D} \right)$ on which constraints on $\left(E/A \right)$ in PNM and $\mn$ act. Runs~1, 1a and 1b.
	}
	\label{Fig:Hist_Prior}
\end{figure*}

The length of the original chain was chosen such that 
(i) the final posterior has at least 100\,000 samples for all runs except run 0, where we were only able to compute $\approx 32\,000$ samples. This run had the highest integrated autocorrelation length ($\tau_\mathrm{orig} \approx 10\,000$) and the lowest acceptance fraction of the post-filtering ($\approx 0.8\%$),
(ii) the length is $\geq 200 \tau_\mathrm{orig}$. This is needed to ensure that the estimation of $\tau_\mathrm{orig}$ is reliable (as mentioned in the \textsc{emcee} documentation, the length of the chain should be at the very least $50 \tau$). 

Apart from the autocorrelation length analysis, we have performed ``bootstrap'' tests of the stability of the posterior distributions of the input parameters to ensure proper convergence of the chains. The procedure is the same as described in our paper \citep{Beznogov_PRC_2023}. The results showed no systematic drifts of the posterior. For run~0 (our worst case run in terms of convergence), we have computed the values of 10\%, 50\% and 90\% quantiles for each bootstrap slice. The maximum difference between corresponding values was $\approx 2\%$. Typical differences were less than 1\%.  

To additionally verify the correctness of the results, the posterior for the most numerically problematic run (run~0) was calculated independently by means of a nested sampling \citep{Skilling_2004,Skilling_2006} algorithm MLFriends \citep{Buchner_2016,Buchner_2019}, implemented in the Python package \textsc{ultranest} (v.3.5.7) \citep{ultranest}\,\footnote{The documentation is available at \url{https://johannesbuchner.github.io/UltraNest/} and the GitHub page is \url{https://github.com/JohannesBuchner/UltraNest}.}. After $\approx 3.1\times 10^9$ likelihood evaluations we ended up with $\approx 90\,000$ effective samples of the posterior (we employed region sampling to avoid any possible biases related to step sampling, see \textsc{ultranest} documentation; this approach might be very inefficient, but robust). Again, we have computed the values of 10\%, 50\% and 90\% quantiles and compared \textsc{ultranest} results against \textsc{emcee} results. And again, the maximum difference between corresponding values was $\approx 2.6\%$ and typical differences were less than 1\%. Plotted on a figure, the histograms were basically identical. Note that in our ``pre-production'' tests we have also utilized extensively Nested Sampling Python package \textsc{dynesty} (v.2.1.0) \citep{dynesty,dynesty2}\,\footnote{The documentation is available at \url{https://dynesty.readthedocs.io/en/stable/} and the GitHub page is \url{https://github.com/joshspeagle/dynesty}.}.

Thus, we have confirmed that despite above-mentioned difficulties and bi-modality of some parameters (e.g., Figs~\ref{Fig:Hist_ModelParam}, \ref{Fig:correl_NM_run0} and \ref{Fig:correl_NM_run1}), our results are reliable. 

Another topic worth discussing in more details is the change of the prior due to re-parametrization. To investigate this question we have computed two simplified analogs of run~1. Run~1a is the full equivalent of run~1 except for prior range of $\sigma$, which was reduced to $[0.1, 1.1]$ from $[0.01, 1.1]$. Since $\sigma = 0$ is a singular point, this change was sufficient to make computation in the effective interactions parameters parametrization feasible. Thus, run~1b is analog of run~1a, but in the effective interaction parameters parametrization. The priors for run~1b were distributed uniformly in the ranges indicated in Table~\ref{tab:Prior-1b}. Those ranges were chosen by analyzing with the results of run~1a. 

The results are presented on Fig.~\ref{Fig:Hist_Prior}, where runs~1, 1a and 1b are confronted against each other. Let us first compare runs~1 and 1a. They use the same ``mixed'' parametrization. We can see that reducing the range of $\sigma$ does not strongly affect the results. $\eff{C}$ and $\eff{D}$	are almost unchanged, while $C_0$, $D_0$, $C_3$ and $D_3$ are much more localized in run~1a. This means that the values of $\sigma$  close to zero are responsible for long tails of $C_0$, $D_0$, $C_3$ and $D_3$ in run~1 (and other runs as well). This is not surprising, as  $\sigma = 0$ is not only a singular point of transformations \eqref{eq:re-param}, but also a degenerate point of energy functionals as discussed in Sec.~\ref{sec:Model}.	Comparing runs~1a and 1b, which differ in parametrization and, thus, in priors, one can conclude that re-parametrization has little impact on the posteriors. Indeed, only the $\sigma$ posterior distribution shows slight difference between the runs, all other parameters are virtually identical. This means that the posterior distributions are strongly dominated by the constraints and are almost insensitive to the priors. This, in turn, justifies our re-parametrization and proves the robustness of the posteriors to the variations of the rather arbitrary chosen priors.

\section{Posterior data tables}
\label{App:Data}

Information complementary to Figs.~\ref{Fig:Hist_NM} and
\ref{Fig:Hist_NS} is provided here.

\renewcommand{\arraystretch}{1.2}
\setlength{\tabcolsep}{3.5pt}
\begin{table*}
    \caption{Parameters of marginalized posterior distributions corresponding to each run. Considered quantities are NEPs; effective masses of nucleons in SMN and PNM at $n=0.16~\mathrm{fm}^{-3}$, respectively; energy per particle in PNM at $n=0.04,~0.08,~0.12$ and $0.16~\mathrm{fm}^{-3}$; key parameters of NSs, as discussed in the text; symmetry energy ($E_{\mathrm{sym};0.1}$) and symmetry pressure ($P_{\mathrm{sym}; 0.1}$) at $n=0.1~\mathrm{fm}^{-3}$. Medians (Med.) and 68\% CI are shown. The median values and the values of the CI are rounded to 3 and 2 significant digits, respectively; \emph{all} trailing zeros after the decimal point are removed.}
	\begin{tabular}{cccccccccccc}
		\toprule
		\toprule
		\multirow{2}{*}{Par.}  &  \multirow{2}{*}{Units}                      & \multicolumn{2}{c}{run 0}                  & \multicolumn{2}{c}{run 1}                  & \multicolumn{2}{c}{run 2}                  & \multicolumn{2}{c}{run 3}                  & \multicolumn{2}{c}{run 4}                  \\
		\cmidrule(lr){3-4}
		\cmidrule(lr){5-6}
		\cmidrule(lr){7-8}
		\cmidrule(lr){9-10}
        \cmidrule(lr){11-12}
                                             &                               &      Med.   &         68\% CI              &      Med.   &            68\% CI           &       Med.  &               68\% CI        &     Med.    &              68\% CI         &      Med.   &        68\% CI               \\ 
        \midrule
        $n_\mathrm{sat}$                     & $\mathrm{fm}^{-3}$            & $     0.16$ & $^{+   0.0039}_{-   0.0039}$ & $     0.16$ & $^{+   0.0039}_{-   0.0039}$ & $     0.16$ & $^{+   0.0038}_{-   0.0039}$ & $     0.16$ & $^{+    0.004}_{-    0.004}$ & $     0.16$ & $^{+   0.0038}_{-   0.0039}$ \\
        $E_\mathrm{sat}$                     & $\mathrm{MeV}$                & $    -15.9$ & $^{+      0.2}_{-      0.2}$ & $    -15.9$ & $^{+      0.2}_{-      0.2}$ & $    -15.9$ & $^{+      0.2}_{-      0.2}$ & $    -15.9$ & $^{+      0.2}_{-      0.2}$ & $    -15.9$ & $^{+      0.2}_{-      0.2}$ \\
        $K_\mathrm{sat}$                     & $\mathrm{MeV}$                & $      265$ & $^{+       33}_{-       25}$ & $      260$ & $^{+       34}_{-       24}$ & $      234$ & $^{+       25}_{-       13}$ & $      297$ & $^{+       10}_{-       17}$ & $      275$ & $^{+       29}_{-       34}$ \\
        $Q_\mathrm{sat}$                     & $\mathrm{MeV}$                & $     -271$ & $^{+       79}_{-       54}$ & $     -283$ & $^{+       78}_{-       51}$ & $     -339$ & $^{+       34}_{-       13}$ & $     -200$ & $^{+       35}_{-       68}$ & $     -259$ & $^{+       84}_{-       69}$ \\
        $Z_\mathrm{sat}$                     & $\mathrm{MeV}$                & $      899$ & $^{+      440}_{-      620}$ & $      990$ & $^{+      420}_{-      620}$ & $     1450$ & $^{+      160}_{-      400}$ & $      315$ & $^{+      360}_{-      190}$ & $      743$ & $^{+      600}_{-      570}$ \\
        $J_\mathrm{sym}$                     & $\mathrm{MeV}$                & $     30.8$ & $^{+     0.99}_{-      1.1}$ & $       31$ & $^{+     0.94}_{-      1.1}$ & $     31.2$ & $^{+     0.78}_{-     0.72}$ & $     30.2$ & $^{+      1.2}_{-      1.8}$ & $     30.7$ & $^{+      1.1}_{-      1.2}$ \\
        $L_\mathrm{sym}$                     & $\mathrm{MeV}$                & $     45.7$ & $^{+      4.9}_{-      5.3}$ & $     51.7$ & $^{+      9.8}_{-      8.4}$ & $     50.7$ & $^{+      6.6}_{-      5.7}$ & $     45.9$ & $^{+      6.9}_{-      8.4}$ & $     48.6$ & $^{+      6.7}_{-      6.5}$ \\
        $K_\mathrm{sym}$                     & $\mathrm{MeV}$                & $     -166$ & $^{+       31}_{-       34}$ & $    -85.4$ & $^{+       79}_{-       87}$ & $    -91.2$ & $^{+       45}_{-       46}$ & $     -181$ & $^{+       30}_{-       24}$ & $     -154$ & $^{+       50}_{-       35}$ \\
        $Q_\mathrm{sym}$                     & $\mathrm{MeV}$                & $      345$ & $^{+       54}_{-       68}$ & $      447$ & $^{+       83}_{-      140}$ & $      463$ & $^{+       56}_{-       76}$ & $      272$ & $^{+       28}_{-       26}$ & $      333$ & $^{+       71}_{-       81}$ \\
        $Z_\mathrm{sym}$                     & $\mathrm{MeV}$                & $    -1930$ & $^{+      530}_{-      440}$ & $-    2860$ & $^{+     1200}_{-      690}$ & $    -2940$ & $^{+      690}_{-      480}$ & $    -1400$ & $^{+      120}_{-      160}$ & $-    1840$ & $^{+      540}_{-      640}$ \\
        $m_{\mathrm{eff;\,N}}^\mathrm{SNM}$  & $m_{\mathrm{N}}$              & $    0.525$ & $^{+     0.12}_{-     0.11}$ & $    0.535$ & $^{+    0.099}_{-    0.098}$ & $    0.637$ & $^{+    0.013}_{-    0.013}$ & $    0.575$ & $^{+     0.19}_{-     0.18}$ & $    0.531$ & $^{+     0.13}_{-     0.18}$ \\
        $(E/A)_1$                            & $\mathrm{MeV}$                & $     6.35$ & $^{+    0.036}_{-    0.034}$ & $     6.62$ & $^{+      0.7}_{-     0.42}$ & $     6.57$ & $^{+     0.39}_{-     0.34}$ & $     6.21$ & $^{+     0.23}_{-     0.25}$ & $     6.17$ & $^{+    0.057}_{-    0.058}$ \\
        $(E/A)_2$                            & $\mathrm{MeV}$                & $     9.41$ & $^{+      0.2}_{-      0.2}$ & $     9.19$ & $^{+     0.22}_{-     0.22}$ & $      9.2$ & $^{+     0.22}_{-     0.22}$ & $      9.1$ & $^{+     0.21}_{-     0.21}$ & $     9.04$ & $^{+     0.13}_{-     0.14}$ \\
        $(E/A)_3$                            & $\mathrm{MeV}$                & $     12.8$ & $^{+     0.46}_{-     0.46}$ & $     12.2$ & $^{+     0.35}_{-     0.37}$ & $     12.2$ & $^{+     0.31}_{-     0.31}$ & $     12.5$ & $^{+     0.28}_{-     0.28}$ & $     12.4$ & $^{+     0.29}_{-     0.29}$ \\
        $(E/A)_4$                            & $\mathrm{MeV}$                & $       17$ & $^{+     0.74}_{-     0.74}$ & $     16.3$ & $^{+     0.64}_{-     0.63}$ & $     16.2$ & $^{+     0.63}_{-     0.62}$ & $     16.7$ & $^{+      0.6}_{-     0.57}$ & $     16.6$ & $^{+     0.54}_{-     0.52}$ \\
        $m_{\mathrm{eff;\,n}}^\mathrm{PNM}$  & $m_{\mathrm{N}}$              & $    0.685$ & $^{+    0.027}_{-    0.039}$ & $    0.522$ & $^{+     0.13}_{-     0.14}$ & $    0.576$ & $^{+    0.078}_{-    0.075}$ & $    0.875$ & $^{+    0.026}_{-    0.026}$ & $    0.628$ & $^{+    0.063}_{-     0.17}$ \\
        $M_\mathrm{G}^*$                     & $\Msun$                       & $     2.02$ & $^{+    0.029}_{-    0.015}$ & $     2.14$ & $^{+     0.14}_{-    0.099}$ & $      2.1$ & $^{+    0.056}_{-     0.06}$ & $     2.06$ & $^{+    0.087}_{-    0.046}$ & $     2.09$ & $^{+     0.07}_{-    0.059}$ \\
        $c^{2}_\mathrm{s}$                   & $c^2$                         & $    0.928$ & $^{+    0.025}_{-    0.028}$ & $    0.948$ & $^{+    0.035}_{-    0.055}$ & $    0.933$ & $^{+    0.043}_{-    0.069}$ & $    0.925$ & $^{+    0.047}_{-    0.054}$ & $     0.95$ & $^{+     0.03}_{-    0.037}$ \\
        $n_\mathrm{c}^*$                     & $\mathrm{fm}^{-3}$            & $     1.21$ & $^{+     0.02}_{-     0.03}$ & $     1.09$ & $^{+      0.1}_{-     0.13}$ & $     1.13$ & $^{+    0.074}_{-    0.059}$ & $     1.16$ & $^{+    0.046}_{-    0.079}$ & $     1.15$ & $^{+    0.056}_{-    0.063}$ \\
        $\rho_\mathrm{c,15}^*$               & $10^{15}~\mathrm{g/cm^{3}}$   & $     2.85$ & $^{+    0.053}_{-    0.072}$ & $     2.57$ & $^{+     0.24}_{-     0.31}$ & $     2.66$ & $^{+     0.19}_{-     0.15}$ & $     2.72$ & $^{+     0.11}_{-     0.18}$ & $      2.7$ & $^{+     0.13}_{-     0.15}$ \\
        $P_\mathrm{c,36}^*$                  & $\mathrm{10^{36}~dyn/cm^{2}}$ & $      1.3$ & $^{+    0.048}_{-    0.054}$ & $      1.2$ & $^{+     0.12}_{-     0.16}$ & $     1.24$ & $^{+      0.1}_{-     0.13}$ & $     1.23$ & $^{+    0.077}_{-    0.069}$ & $     1.26$ & $^{+    0.057}_{-    0.058}$ \\
        $R_{1.4}$                            & $\mathrm{km}$                 & $     11.8$ & $^{+     0.14}_{-     0.12}$ & $     12.2$ & $^{+     0.61}_{-     0.41}$ & $     12.1$ & $^{+     0.31}_{-     0.31}$ & $       12$ & $^{+     0.24}_{-      0.2}$ & $     12.1$ & $^{+     0.26}_{-     0.22}$ \\
        $\Lambda_{1.4}$                      & $-$                           & $      315$ & $^{+       27}_{-       20}$ & $      408$ & $^{+      170}_{-       86}$ & $      372$ & $^{+       71}_{-       61}$ & $      361$ & $^{+       56}_{-       34}$ & $      363$ & $^{+       56}_{-       44}$ \\
        $R_{2.0}$                            & $\mathrm{km}$                 & $     10.6$ & $^{+      0.3}_{-     0.21}$ & $     11.6$ & $^{+     0.93}_{-     0.79}$ & $     11.3$ & $^{+     0.44}_{-     0.59}$ & $     11.1$ & $^{+     0.57}_{-     0.46}$ & $     11.2$ & $^{+     0.47}_{-     0.51}$ \\
        $\Lambda_{2.0}$                      &  $-$                          & $     10.1$ & $^{+      3.1}_{-      1.8}$ & $     23.6$ & $^{+       22}_{-       11}$ & $       19$ & $^{+      7.6}_{-      7.5}$ & $     15.4$ & $^{+      9.2}_{-      5.2}$ & $     17.1$ & $^{+      7.6}_{-      6.1}$ \\
        $E_{\mathrm{sym};0.1}$               & $\mathrm{MeV} $               &      $23.6$ & $^{+      0.6}_{-      0.7}$ &      $23.7$ & $^{+      0.5}_{-      0.6}$ &        $24$ & $^{+     0.4 }_{-    0.4  }$ &      $22.9$ & $^{+      0.6}_{-        1}$ &      $23.3$ & $^{+      0.6}_{-      0.8}$ \\
        $P_{\mathrm{sym};0.1}$               & $\mathrm{MeV/fm^{3}}$         &      $1.45$ & $^{+     0.08}_{-     0.07}$ &       $1.4$ & $^{+      0.1}_{-     0.09}$ &       $1.4$ & $^{+      0.1}_{-     0.09}$ &       $1.5$ & $^{+      0.08}_{-    0.14}$ &       $1.5$ & $^{+     0.09}_{-     0.08}$ \\
		\bottomrule
		\bottomrule
	\end{tabular}
	\label{tab:Posteriors}
\end{table*}
\renewcommand{\arraystretch}{1.0}
\setlength{\tabcolsep}{2.0pt}

\bibliographystyle{aasjournal-hyperref}
\bibliography{Skyrme.bib,MCMC.bib}

\end{document}